\documentclass[11pt,a4paper]{article}
\usepackage[english]{babel}
\usepackage{amsmath,amsthm,amssymb,epsfig,latexsym}
\usepackage{color}
\usepackage{ulem}
\usepackage[numbers,square,sort&compress]{natbib}

\newcommand{\so}{\scriptscriptstyle \rm I}
\newcommand{\st}{\scriptscriptstyle \rm I\hspace{-1pt}I}
\newcommand{\sth}{\scriptscriptstyle \rm I\hspace{-1pt}I\hspace{-1pt}I}
\newcommand{\qo}{\rm i}
\newcommand{\qt}{\rm ii}
\newcommand{\qth}{\rm iii}

\newcommand{\bu}{\bar u}
\newcommand{\bv}{\bar v}

\newcommand{\bt}{\bar t}
\newcommand{\bx}{\bar x}
\newcommand{\bs}{\bar s}
\newcommand{\by}{\bar y}
\newcommand{\bz}{\bar z}

\newcommand{\vc}{v^{\scriptscriptstyle C}}
\newcommand{\vb}{v^{\scriptscriptstyle B}}
\newcommand{\buc}{\bar{u}^{\scriptscriptstyle C}}
\newcommand{\bub}{\bar{u}^{\scriptscriptstyle B}}
\newcommand{\bvc}{\bar{v}^{\scriptscriptstyle C}}
\newcommand{\bvb}{\bar{v}^{\scriptscriptstyle B}}

\newcommand{\bet}{\bar\eta}

\newcommand{\bw}{\bar w}

\newcommand{\bbet}{\bar\beta}
\newcommand{\bxi}{\bar\xi}

\newcommand{\be}[1]{\begin{equation}\label{#1}}
\newcommand{\ba}[1]{\begin{multline}\label{#1}}
\newcommand{\ee}{\end{equation}}
\newcommand{\ea}{\end{eqnarray}}

\newcommand{\num}{\\\rule{0pt}{20pt}}

\newcommand{\str}{\mathop{\rm str}}

\newcommand{\mb}[1]{\quad\mbox{#1}\quad}
\newcommand{\wh}[1]{\widehat{#1}}

\newtheorem{prop}{Proposition}[section]
\newtheorem{lemma}{Lemma}[section]

\newtheorem{Def}{Definition}[section]

\newtheorem{rmk}{Remark}[section]
\def\qed{\hfill\nobreak\hbox{$\square$}\par\medbreak}
 \makeatletter
 \@addtoreset{equation}{section}
 \makeatother

\newcommand{\bea}{\begin{eqnarray}}
\newcommand{\eea}{\end{eqnarray}}

\newcommand{\bT}{\mathbb{T}}

\begin{document}

\begin{flushright}
LAPTH-024/16
\end{flushright}

\vspace{16pt}

\begin{center}
\begin{LARGE}
{\bf Scalar products of Bethe vectors \\[2mm] in models with $\mathfrak{gl}(2|1)$ symmetry\\[2mm]
 1. Super-analog of Reshetikhin formula}
\end{LARGE}

\vspace{40pt}

\begin{large}
{A.~Hutsalyuk${}^{a}$,  A.~Liashyk${}^{b,c}$,
S.~Z.~Pakuliak${}^{a,d}$,\\ E.~Ragoucy${}^e$, N.~A.~Slavnov${}^f$  \footnote{%
hutsalyuk@gmail.com, a.liashyk@gmail.com, stanislav.pakuliak@jinr.ru, eric.ragoucy@lapth.cnrs.fr,\\ nslavnov@mi.ras.ru}}
\end{large}

 \vspace{12mm}

${}^a$ {\it Moscow Institute of Physics and Technology,  Dolgoprudny, Moscow reg., Russia}

\vspace{4mm}

${}^b$ {\it Bogoliubov Institute for Theoretical Physics, NAS of Ukraine,  Kiev, Ukraine}

\vspace{4mm}

${}^c$ {\it National Research University Higher School of Economics,  Russia}

\vspace{4mm}

${}^d$ {\it Laboratory of Theoretical Physics, JINR,  Dubna, Moscow reg., Russia}

\vspace{4mm}

${}^e$ {\it Laboratoire de Physique Th\'eorique LAPTH, CNRS and Universit\'e de Savoie,\\
BP 110, 74941 Annecy-le-Vieux Cedex, France}

\vspace{4mm}

${}^f$ {\it Steklov Mathematical Institute of Russian Academy of Sciences, Moscow, Russia}

\end{center}

%%%%%%%%%%%%%%%%%%%%%%%%%%%%%%%

\vspace{1cm}

\centerline{ Dedicated to the memory of P.P. Kulish}

\vspace{4mm}

\begin{abstract}
We study  scalar products of Bethe vectors in integrable models solvable by nested algebraic Bethe ansatz and
possessing $\mathfrak{gl}(2|1)$ symmetry. Using explicit formulas  of the monodromy matrix entries multiple actions
onto Bethe vectors we obtain a representation for the scalar product in the most general case. This explicit representation
appears to be a sum over partitions of the Bethe parameters. It can be used for the analysis of scalar products involving
on-shell Bethe vectors.

 As a by-product, we obtain a determinant representation for the scalar products of generic Bethe vectors in integrable models
with $\mathfrak{gl}(1|1)$ symmetry.
\end{abstract}

\vspace{1cm}
%\centerline{Revised {\red \today}}

\section{Introduction}

 The study of quantum integrable systems in the context of Quantum Inverse Scattering Method (QISM) has been one of the great successes of the Leningrad school \cite{FadST79,FadT79,FadLH96}. As one of the early members of this school, Petr Kulish has made several important contributions to the field. Among them, it is worth noticing the study of quantum integrable models solvable by the nested algebraic Bethe ansatz \cite{Kul81,KulRes81,KulRes83}, the search for solutions to the Yang-Baxter equation \cite{KulS80,KulRes82}, or the development of QISM for models with open boundaries \cite{KulSas93,Damk03,AvaKR11}. He also contributed to the generalisations to the supersymmetric versions of these models \cite{KulS80,Kul85}.

In this article, we would like to make a contribution to the field he was found of, that is supersymmetric quantum integrable models solvable  by the nested algebraic Bethe ansatz. This paper is the continuation of a series of articles devoted to quantum integrable models
possessing a $\mathfrak{gl}(2|1)$ symmetry.
  In \cite{PakRS16a} we have  built explicit representations for the Bethe vectors in these models.
In \cite{HutLPRS16a} we found multiple actions of the monodromy matrix entries onto these vectors. Now we consider the problem of scalar products
of Bethe vectors.

Scalar products of Bethe vectors play a very important role in the Algebraic Bethe ansatz. They are a necessary tool for
calculating form factors and correlation functions within this framework. The first results in this field concern
 $\mathfrak{gl}(2)$-based models and their $q$-deformations. They were  obtained in \cite{Kor82,IzeK84,Ize87}, where in particular the Izergin--Korepin formula (for scalar products) was given. Concerning scalar
products in models with $\mathfrak{gl}(3)$-invariant
$R$-matrix, the first result was obtained  in \cite{Res86}. There, an analog of the Izergin--Korepin
formula  for scalar product of generic Bethe vectors was produced (the Reshetikhin formula) and a determinant representation
for the norm of the eigenvectors of the transfer matrix was found. Similar results for the models based on a $q$-deformed $\mathfrak{gl}(3)$ algebra
were obtained in \cite{PakRS14a,Sla15a}.

In this paper we consider scalar products in the models with  $\mathfrak{gl}(2|1)$ symmetry. At this stage of study our goal is to obtain an analog
of Reshetikhin formula for these models. This means that we are going to find an explicit representation for the scalar product
of generic Bethe vectors, in which the result is given as a sum over partitions of the Bethe parameters (sum formula). Certainly, this type
of formulas is not convenient for direct applications. However, they give a key for studying particular cases of scalar product, in which the sum over partitions can be reduced to a single determinant. Such determinant representations
for particular cases of scalar products were already found for the models with $\mathfrak{gl}(2)$ and $\mathfrak{gl}(3)$ symmetries
\cite{Sla89,KitMT99,BelPRS12b,Sla15a}. In all these cases the starting point was a sum formula.

The article is organized as follows. In section~\ref{S-N} we introduce the model under consideration and
specify our conventions and notation. Section~\ref{S-NT} contains explicit formulas for the  Bethe  vectors and multiple
actions of the monodromy matrix entries onto them.
In section~\ref{S-GFSP} we present the main result of the paper. There we define  scalar products
and give  a sum formula for them. The sum formula is proved in the remaining part of the paper.
In section~\ref{S-SA} we study successive actions of the monodromy matrix entries onto Bethe vectors.
In section~\ref{S-HC} we  obtain an explicit determinant representation for the highest coefficient of the scalar product.
Section~\ref{S-SP11} deals with a particular case of the scalar products in the $\mathfrak{gl}(1|1)$-based models.
Section~\ref{S-DRHC} contains several alternative representations for the highest coefficient.
We have collected auxiliary formulas in four appendices. In particular, appendix~\ref{A-IHC} is devoted to the properties of
the partition function of six-vertex model with domain wall boundary conditions. In appendix~\ref{A-SumInt} we describe a relationship of sums over
partitions and multiple contour integrals. Appendix~\ref{A-SF} contains several identities allowing us to calculate certain multiple sums of rational functions.
Finally, in appendix~\ref{A-RPJ} we present reduction properties of a determinant introduced in section~\ref{S-HC}.

\section{Description of the model\label{S-N}}

\subsection{$\mathfrak{gl}(2|1)$-based models}

The $R$-matrix of $\mathfrak{gl}(2|1)$-based models acts in the tensor product $\mathbb{C}^{2|1}\otimes \mathbb{C}^{2|1}$,
where $\mathbb{C}^{2|1}$ is the $\mathbb{Z}_2$-graded vector space with the grading $[1]=[2]=0$, $[3]=1$.
%We refer the reader
%for more detailed description of the algebraic Bethe ansatz for models with grading to the works \cite{aaa}
The $R$-matrix has the form
 \be{R-mat}
 R(u,v)=\mathbb{I}+g(u,v)P, \qquad g(u,v)=\frac{c}{u-v},
 \ee
where $\mathbb{I}$ is the identity matrix, $P$ is the graded permutation operator \cite{KulS80} and $c$ is a constant. The
monodromy matrix $T(u)$ is also graded according to the rule $[T_{ij}(u)]=[i]+[j]$. It satisfies the $RTT$-relation
\be{RTT}
R(u,v)\bigl(T(u)\otimes \mathbb{I}\bigr) \bigl(\mathbb{I}\otimes T(v)\bigr)= \bigl(\mathbb{I}\otimes T(v)\bigr)
\bigl( T(u)\otimes \mathbb{I}\bigr)R(u,v).
\ee
Equation \eqref{RTT} holds in the tensor product $\mathbb{C}^{2|1}\otimes \mathbb{C}^{2|1}\otimes\mathcal{H}$,
where $\mathcal{H}$ is a Hilbert space of a Hamiltonian  under consideration. Tensor products
of $\mathbb{C}^{2|1}$ spaces are graded as follows:
\be{tens-prod}
(\mathbb{I}\otimes e_{ij})\,\cdot\,(e_{kl}\otimes \mathbb{I}) = (-1)^{([i]+[j])([k]+[l])}\,e_{kl}\otimes e_{ij},
\ee
where $e_{ij}$ are the elementary units: $(e_{ij})_{ab}=\delta_{ia}\delta_{jb}$.

Algebra \eqref{RTT} possesses an antimorphism
\cite{PakRS16a}
\be{psi}
\psi\bigl(T_{ij}(u)\bigr)=(-1)^{[i][j]+[i]}T_{ji}(u),\qquad
\psi\bigl(AB\bigr)=(-1)^{[A][B]}\psi\bigl(B\bigr)\psi\bigl(A\bigr),
\ee
where $A$ and $B$ are arbitrary operators of fixed gradings. It follows from \eqref{psi} that
\be{psiAn}
\psi\bigl(A_1\dots A_n\bigr)=(-1)^{\vartheta_n}\psi\bigl(A_n\bigr)\dots\psi\bigl(A_1\bigr), \qquad
\vartheta_n=\sum_{1\le i<j\le n} [A_i]\cdot [A_j].
\ee

The $RTT$-relation \eqref{RTT} implies a set of scalar commutation relations for the monodromy matrix elements.
 For our purpose, we will need only the following ones:
\be{odd-comm}
\begin{aligned}
&[T_{ik}(v_1),T_{ik}(v_2)]=0,\qquad \text{if}\quad [T_{ik}]=0,\\
& h(v_1,v_2)T_{j3}(v_1)T_{j3}(v_2)=h(v_2,v_1)T_{j3}(v_2)T_{j3}(v_1),\qquad j=1,2.\\
&h(v_2,v_1)T_{3j}(v_1)T_{3j}(v_2)=h(v_1,v_2)T_{3j}(v_2)T_{3j}(v_1),\qquad j=1,2.
\end{aligned}
\ee

Finally, the graded transfer matrix is defined as the supertrace of the monodromy matrix
\be{transfer.mat}
\mathcal{T}(u)=\str T(u)= \sum_{j=1}^{3} (-1)^{[j]}\, T_{jj}(u).
\ee
It is a generating function of the integrals of motion, due to the relation $[\mathcal{T}(u)\,,\,\mathcal{T}(v)]=0$.

\subsection{Notation and known results}

In this paper we use the same notation and conventions as in \cite{HutLPRS16a}. Besides the function $g(u,v)$
we shall use also two functions
\be{desand}
f(u,v)=1+g(u,v)=\frac{u-v+c}{u-v} \mb{and}
h(u,v)=\frac{f(u,v)}{g(u,v)}=\frac{u-v+c}{c}.
\ee
These functions have the following obvious properties:
 \be{propert}
 g(u,v)=-g(v,u),\quad h(u,v+c)=\frac1{g(u,v)},\quad  f(u,v+c)=\frac1{f(v,u)}.
 \ee

We  denote sets of variables by bar: $\bx$, $\bu$, $\bv$ etc.
Individual elements of the sets are denoted by Latin subscripts: $v_j$, $u_k$ etc. As a rule, the number of elements in the
sets is not shown explicitly in the equations, however we give these cardinalities in
special comments to the formulas. The notation $\bu\pm c$ means that all the elements of the set $\bu$ are
shifted by $\pm c$:  $\bu\pm c=\{u_1\pm c,\dots,u_n\pm c\}$. A union of sets is denoted by braces: $\{\bu,\bv\}\equiv
\bu\cup \bv$.

As a rule, subsets of variables are labeled by Roman subscripts: $\bu_{\so}$, $\bv_{\rm ii}$, $\bx_{\st}$ etc.  The only exception
will be the notation $\bu_j$ for a subset that is complementary to the element $u_j$, that is $\bu_j=\bu\setminus \{u_j\}$.
A notation $\bu\Rightarrow\{\bu_{\so},\bu_{\st}\}$ means that the
set $\bu$ is divided into two  subsets $\bu_{\so}$ and $\bu_{\st}$ such that $\{\bu_{\so},\bu_{\st}\}=\bu$ and  $\bu_{\so}\cap\bu_{\st}
=\emptyset$. We assume that the elements in every subset of variables are ordered in such a way that the sequence of
their subscripts is strictly increasing. We call this ordering  natural order.

In order to avoid too cumbersome formulas we use a shorthand notation for products of  functions depending on one or two variables.
 Namely, if  the functions $g$, $f$, $h$ depend
on a set of variables, this means that one should take the product over the corresponding set.
For example,
 \be{SH-prodllll}
 h(\bu,v)= \prod_{u_j\in\bu} h(u_j,v);\quad
 f(u_k,\bv_j)=\prod_{\substack{v_l\in\bv\\ l\ne j}} f(u_k,v_l);\quad
  g(\bv_{\so},\bv_{\st})=\prod_{v_j\in\bv_{\so}}\prod_{v_k\in\bv_{\st}} g(v_j,v_k).
 \ee
We use the same prescription for the products of even commuting operators $T_{ij}$ and for vacuum eigenvalues
$\lambda_k$ and $r_k$ (see \eqref{Tjj}, \eqref{ratios}), for instance,
 \be{SH-prodlllo}
  \lambda_2(\bv)= \prod_{v_j\in\bv} \lambda_2(v_j);\quad  r_3(\bu)= \prod_{u_j\in\bu} r_3(u_j);\quad  T_{21}(\bv_{\st})= \prod_{v_j\in \bv_{\st}} T_{21}(v_j).
 \ee

For the odd operators $T_{j3}$ and $T_{3j}$, that exchange with a multiplication factor,  we introduce for an arbitrary set
$\bv=\{v_1,\dots,v_n\}$
\be{bTc-def}
\bT_{j3}(\bv)= \frac{T_{j3}(v_1)\dots T_{j3}(v_n)}{\prod_{n\ge \ell>m\ge 1} h(v_\ell,v_m)}, \qquad
\bT_{3j}(\bv)= \frac{T_{3j}(v_1)\dots T_{3j}(v_n)}{\prod_{n\ge \ell>m\ge 1} h(v_m,v_\ell)}, \qquad j=1,2.
\ee
 Due to commutation relations \eqref{odd-comm} the operator products \eqref{bTc-def} are symmetric over
the parameters $\bv$ and play the same role as the bosonic products $T_{12}(\bar u)$.

We would like to stress that the convention on the shorthand notation for the products concerns the functions (operators)
depending on one or two variables only. It should not be applied to the functions, which by definition might depend
on many variables.

An example of a function depending on two sets of variables
is the partition function of the six-vertex model with domain wall boundary conditions (DWPF) \cite{Kor82,Ize87}.  We denote it by
$K_n(\bu|\bv)$. It depends on two sets of variables $\bu$ and $\bv$, and the subscript shows that
$\#\bu=\#\bv=n$. The function $K_n$ has the following determinant representation
\begin{equation}\label{K-def}
K_n(\bu|\bv)
=\Delta'_n(\bu)\Delta_n(\bv)h(\bu,\bv)
\det_n \left(\frac{g(u_j,v_k)}{h(u_j,v_k)}\right),
\end{equation}
where $\Delta'_n(\bu)$ and $\Delta_n(\bv)$ are
\be{def-Del}
\Delta'_n(\bu)
=\prod_{j<k}^n g(u_j,u_k),\qquad {\Delta}_n(\bv)=\prod_{j>k}^n g(v_j,v_k).
\ee
It is easy to see that $K_n$ is symmetric over $\bu$ and symmetric over $\bv$, however  $K_n(\bu|\bv)\ne
 K_n(\bv|\bu)$. Some other properties of DWPF are given in appendix~\ref{A-IHC}.

\section{Necessary tools\label{S-NT}}

In this section we recall some explicit formulas for the Bethe vectors and describe the multiple actions
of the operators $T_{ij}$ onto them.

\subsection{Bethe vectors\label{S-BV}}

Generic Bethe vectors and their dual vectors are denoted respectively by $\mathbb{B}_{a,b}(\bu;\bv)$
and $\mathbb{C}_{a,b}(\bu;\bv)$. They are parameterized by two sets of
complex parameters (Bethe parameters) $\bu=u_1,\dots,u_a$ and $\bv=v_1,\dots,v_b$ with $a,b=0,1,\dots$. The reader
can find several explicit representations for the Bethe vectors in terms of the monodromy matrix entries in \cite{PakRS16a}.
 Here we use some of them (see below). Vectors $|0\rangle=\mathbb{B}_{0,0}(\emptyset;\emptyset)$ and $\langle0|=\mathbb{C}_{0,0}(\emptyset;\emptyset)$
respectively are called a pseudovacuum vector  and a dual pseudovacuum vector. They
are eigenvectors of the diagonal entries of the monodromy matrix
 \be{Tjj}
 T_{ii}(u)|0\rangle=\lambda_i(u)|0\rangle, \qquad   \langle0|T_{ii}(u)=\lambda_i(u)\langle0|,
 \qquad i=1,2,3,
 \ee
where $\lambda_i(u)$ are some scalar functions. In the framework of the generalized model \cite{Kor82} considered in this paper, they remain free functional parameters.
Below it will be convenient to deal with ratios of these functions
 \be{ratios}
 r_1(u)=\frac{\lambda_1(u)}{\lambda_2(u)}, \qquad  r_3(u)=\frac{\lambda_3(u)}{\lambda_2(u)}.
 \ee

Now let us give one of explicit representations for Bethe vectors obtained in \cite{PakRS16a,Sla16}
\be{Phi-expl1}
\mathbb{B}_{a,b}(\bu;\bv)=\sum K_n(\bv_{\so}|\bu_{\so})\frac{ f(\bu_{\so},\bu_{\st}) g(\bv_{\st},\bv_{\so})}
{\lambda_2(\bv)\lambda_2(\bu_{\st})f(\bv,\bu)} \bT_{13}(\bv_{\so})\,\bT_{23}(\bv_{\st})\,T_{12}(\bu_{\st})|0\rangle.
\ee
Here $K_n(\bv_{\so}|\bu_{\so})$ is the DWPF \eqref{K-def}. The sum is taken over partitions $\bu\Rightarrow\{\bu_{\so},\bu_{\st}\}$ and $\bv\Rightarrow\{\bv_{\so},\bv_{\st}\}$,
where $\#\bv_{\so}=\#\bu_{\so}=n$, and $n=0,1,\dots,\min(a,b)$. Recall that the notation $T_{12}(\bu_{\st})$, $ g(\bv_{\st},\bv_{\so})$,
and so on means the products of the operators (functions) over the corresponding subset (see \eqref{SH-prodllll}--\eqref{bTc-def}).

Explicit representations for dual Bethe vectors will play more important role in our calculations.  We use two of them \cite{PakRS16a,Sla16}:
\be{dPhi-expl1}
\mathbb{C}_{a,b}(\bu;\bv)=(-1)^{\frac{b^2-b}2}\sum K_n(\bv_{\so}|\bu_{\so})\frac{ f(\bu_{\so},\bu_{\st}) g(\bv_{\st},\bv_{\so})}
{\lambda_2(\bv)\lambda_2(\bu_{\st})f(\bv,\bu)} \langle0|T_{21}(\bu_{\st})\,\bT_{32}(\bv_{\st})\bT_{31}(\bv_{\so}),
\ee
and
\be{dPhi-expl2}
\mathbb{C}_{a,b}(\bu;\bv)=(-1)^{\frac{b^2-b}2}\sum g(\bv_{\so},\bu_{\so}) \frac{ f(\bu_{\st},\bu_{\so})f(\bv_{\so},\bu_{\st}) g(\bv_{\st},\bv_{\so})h(\bu_{\so},\bu_{\so})}
{\lambda_2(\bu)\lambda_2(\bv_{\st})f(\bv,\bu)}\;
\langle0|\bT_{32}(\bv_{\st})\,\bT_{31}(\bu_{\so})\, T_{21}(\bu_{\st}).
\ee
Here the sum is taken over the same partitions of the sets $\bu$ and $\bv$ as in \eqref{Phi-expl1}.

\subsection{Multiple actions of $T_{ij}$.\label{S-act}}

Explicit formulas for the multiple actions of the operators $T_{ij}$ onto Bethe vectors were obtained
in \cite{HutLPRS16a}. For our goal we need the actions of  $T_{ij}$ with $i>j$. Below we give the corresponding formulas.
Everywhere in this section $\bet=\{\bu,\bar z\}$, $\bxi=\{\bv,\bar z\}$, $\#\bu=a$, $\#\bv=b$, and $\#\bar z=n$.

\subsubsection{Multiple action of $T_{21}$.\label{S-act21}}
The multiple action of the operators $T_{21}$ onto Bethe vector reads
\begin{multline}\label{A-T21}
T_{21}(\bar z)\mathbb{B}_{a,b}(\bu;\bv)=\lambda_2(\bar z)h(\bxi,\bar z)\sum r_1(\bet_{\so})
\frac{f(\bet_{\st},\bet_{\so})f(\bet_{\st},\bet_{\sth})f(\bet_{\sth},\bet_{\so})g(\bxi_{\st},\bxi_{\so})}
{h(\bxi_{\so},\bar z)f(\bxi_{\st},\bet_{\so})}\\
\times K_n(\bar z|\bet_{\st}+c)K_n(\bet_{\so}|\bxi_{\so}+c)
\mathbb{B}_{a-n,b}(\bet_{\sth};\bxi_{\st}).
\end{multline}
Here the functions $K_n$ are the DWPF \eqref{K-def}. The sum is taken over partitions $\bxi\Rightarrow\{\bxi_{\so},\bxi_{\st}\}$ and
$\bet\Rightarrow\{\bet_{\so},\bet_{\st},\bet_{\sth}\}$ with $\#\bxi_{\so}=\#\bet_{\so}=\#\bet_{\st}=n$.
If $n>a$, then the product $T_{21}(\bar z)$ annihilates $\mathbb{B}_{a,b}(\bu;\bv)$.

\subsubsection{Multiple action of $T_{31}$.\label{S-act31}}

The multiple action of $T_{31}(z)$ has the following form:
\begin{multline}\label{A-T310}
\bT_{31}(\bar z)\mathbb{B}_{a,b}(\bu;\bv)=(-1)^{\frac{n(n+1)}2}\lambda_2(\bar z)h(\bxi,\bar z)\sum r_3(\bxi_{\so})r_1(\bet_{\st})
\frac{g(\bxi_{\st},\bxi_{\so})g(\bxi_{\sth},\bxi_{\st})g(\bxi_{\sth},\bxi_{\so})}
{h(\bet_{\so},\bar z)h(\bxi_{\so},\bet_{\so})h(\bxi_{\st},\bar z)}\\
\times \frac{f(\bet_{\so},\bet_{\st}) f(\bet_{\so},\bet_{\sth}) f(\bet_{\sth},\bet_{\st})          h(\bet_{\so},\bet_{\so})}{f(\bxi_{\so},\bet_{\st})f(\bxi_{\so},\bet_{\sth})f(\bxi_{\sth},\bet_{\st})}
\;K_n(\bet_{\st}|\bxi_{\st}+c)\;\mathbb{B}_{a-n,b-n}(\bet_{\sth};\bxi_{\sth}).
\end{multline}
Here the sum is taken over partitions $\bxi\Rightarrow\{\bxi_{\so},\bxi_{\st},\bxi_{\sth}\}$ and
$\bet\Rightarrow\{\bet_{\so},\bet_{\st},\bet_{\sth}\}$ with $\#\bxi_{\so}=\#\bxi_{\st}=\#\bet_{\so}=\#\bet_{\st}=n$.
If $n>\min(a,b)$, then the product $\bT_{31}(\bar z)$ annihilates $\mathbb{B}_{a,b}(\bu;\bv)$.

Actually, for this action  we will need only the particular case $n=a$. Then
\begin{multline}\label{A-T31}
\bT_{31}(\bar z)\mathbb{B}_{a,b}(\bu;\bv)=(-1)^{\frac{n(n+1)}2}\lambda_2(\bar z)h(\bxi,\bar z)\sum r_3(\bxi_{\so})r_1(\bet_{\st})
\frac{g(\bxi_{\st},\bxi_{\so})g(\bxi_{\sth},\bxi_{\st})g(\bxi_{\sth},\bxi_{\so})}
{h(\bet_{\so},\bar z)h(\bxi_{\so},\bet_{\so})h(\bxi_{\st},\bar z)}\\
\times \frac{f(\bet_{\so},\bet_{\st})
h(\bet_{\so},\bet_{\so})}{f(\bxi_{\so},\bet_{\st})f(\bxi_{\sth},\bet_{\st})}
\;K_n(\bet_{\st}|\bxi_{\st}+c)\;\mathbb{B}_{0,b-n}(\emptyset;\bxi_{\sth}).
\end{multline}
Here the sum is taken over partitions $\bxi\Rightarrow\{\bxi_{\so},\bxi_{\st},\bxi_{\sth}\}$ and
$\bet\Rightarrow\{\bet_{\so},\bet_{\st}\}$ with $\#\bxi_{\so}=\#\bxi_{\st}=\#\bet_{\so}=\#\bet_{\st}=n$.

\subsubsection{Multiple action of $T_{32}$.\label{S-act32}}

The multiple action of $T_{32}(z)$ reads
\begin{multline}\label{A-T32}
\bT_{32}(\bar z)\mathbb{B}_{a,b}(\bu;\bv)=(-1)^{\frac{n(n-1)}2}\lambda_2(\bar z)h(\bxi,\bar z)\sum r_3(\bxi_{\so})
\frac{f(\bet_{\so},\bet_{\st}) g(\bxi_{\st},\bxi_{\so})g(\bxi_{\sth},\bxi_{\st})g(\bxi_{\sth},\bxi_{\so})}
{h(\bet_{\so},\bar z)h(\bxi_{\so},\bet_{\so})h(\bxi_{\st},\bar z)f(\bxi_{\so},\bet_{\st})}\\
\times h(\bet_{\so},\bet_{\so})\;\mathbb{B}_{a,b-n}(\bet_{\st};\bxi_{\sth}).
\end{multline}
Here the sum is taken over partitions $\bxi\Rightarrow\{\bxi_{\so},\bxi_{\st},\bxi_{\sth}\}$ and
$\bet\Rightarrow\{\bet_{\so},\bet_{\st}\}$ with $\#\bxi_{\so}=\#\bxi_{\st}=\#\bet_{\so}=n$. If $n>b$,
then the result of this action vanishes.

If $a=0$ and $n=b$, then
\begin{equation}\label{A-T32b2}
\bT_{32}(\bar z)\mathbb{B}_{0,b}(\emptyset;\bv)=(-1)^{\frac{b(b-1)}2}\lambda_2(\bar z)\sum r_3(\bxi_{\so})
g(\bxi_{\st},\bxi_{\so})|0\rangle.
\end{equation}
The sum is taken over partitions $\bxi\Rightarrow\{\bxi_{\so},\bxi_{\st}\}$  with $\#\bxi_{\so}=\#\bxi_{\st}=n$.

\section{General form of the scalar product\label{S-GFSP}}

The scalar product of Bethe vectors is defined as
\be{Def-SP}
S_{a,b}\equiv S_{a,b}(\buc;\bvc|\bub;\bvb)=\mathbb{C}_{a,b}(\buc;\bvc)\mathbb{B}_{a,b}(\bub;\bvb)\,,
\ee
where all the Bethe parameters are generic complex numbers. We have added the superscripts $C$ and $B$
to the sets $\bu$, $\bv$ in order to stress that the vectors
$\mathbb{C}_{a,b}$ and $\mathbb{B}_{a,b}$ may depend on different sets of parameters.

Being a scalar function, the scalar product is invariant under the action of the antimorphism $\psi$ \eqref{psi}
\be{psiSP1}
\psi\Bigl(S_{a,b}(\buc;\bvc|\bub;\bvb)\Bigr)=S_{a,b}(\buc;\bvc|\bub;\bvb).
\ee
On the other hand, acting with $\psi$ on the r.h.s. of \eqref{Def-SP} and using the explicit representations
\eqref{Phi-expl1} and \eqref{dPhi-expl1} for the Bethe vectors we find
\be{psiSP2}
\psi\Bigl(\mathbb{C}_{a,b}(\buc;\bvc)\mathbb{B}_{a,b}(\bub;\bvb)\Bigr)=\mathbb{C}_{a,b}(\bub;\bvb)\mathbb{B}_{a,b}(\buc;\bvc)=S_{a,b}(\bub;\bvb|\buc;\bvc).
\ee
Here we have used $\psi(T_{j3})=T_{3j}$ and $\psi(T_{3j})=-T_{j3}$ for $j=1,2$ (see \eqref{psi}). Then using \eqref{psiAn}, and the fact that the total
number of odd operators $T_{3j}$ and $T_{j3}$ with $j=1,2$ in the scalar product is equal $2b$, we arrive at \eqref{psiSP2}.
Thus, we conclude that the scalar product is invariant under the permutation of the sets
$\{\buc,\bvc\}\leftrightarrow\{\bub,\bvb\}$:
\be{SP-CB}
S_{a,b}(\buc;\bvc|\bub;\bvb)=S_{a,b}(\bub;\bvb|\buc;\bvc).
\ee

In order to calculate the scalar product one can take an explicit formula for the dual Bethe vector (\eqref{dPhi-expl1} or \eqref{dPhi-expl2})
and then use the  formulas of the multiple actions \eqref{A-T21}--\eqref{A-T32b2}.
Basing on  these formulas we can present
the scalar product of Bethe vectors  in the following schematic form:
\be{Sab-genform}
S_{a,b}(\buc;\bvc|\bub;\bvb)=\sum r_1(\bw_{\qo})r_3(\bw_{\qt})W_{\text{part}}(\bw_{\qo};\bw_{\qt};\bw_{\qth}).
\ee
Here a set $\bw$ is the union of all the Bethe parameters: $\bw=\{\buc,\bub,\bvc,\bvb\}$. The sum is taken over partitions
of this set into three subsets $\bw\Rightarrow\{\bw_{\qo},\bw_{\qt},\bw_{\qth}\}$. The functions $W_{\text{part}}$ are some
rational coefficient. Their explicit forms are not important for now. We stress in \eqref{Sab-genform} that a part of the Bethe parameters $\bw_{\qo}$ becomes the arguments of the functions $r_1$, while the parameters $\bw_{\qt}$ become the arguments of the functions $r_3$.  The remaining
parameters $\bw_{\qth}$ enter the rational functions $W_{\text{part}}$ only.

Let us call the set $\{\buc,\bub\}$ the parameters of $u$-type. Correspondingly, we call the set $\{\bvc,\bvb\}$
the parameters of $v$-type.

\begin{prop}\label{uv-type}
The set $\bw_{\qo}$ in \eqref{Sab-genform} consists of parameters of the $u$-type only, while the set $\bw_{\qt}$
consists of the parameters of $v$-type, that is, $\bw_{\qo}\subset \{\buc,\bub\}$ and $\bw_{\qt}\subset \{\bvc,\bvb\}$.
Moreover, $\#\bw_{\qo}=a$ and $\#\bw_{\qt}=b$.
\end{prop}

{\sl Proof.} Let us prove that $\bw_{\qo}\subset \{\buc,\bub\}$. For this we take the dual Bethe vector in the form \eqref{dPhi-expl2}.
Let us fix a partition $\buc\Rightarrow\{\buc_{\so},\buc_{\st}\}$ in \eqref{dPhi-expl2}, such that $\#\buc_{\so}=n$, $n=0,1,\dots,\min(a,b)$.
Calculating the scalar product we first act with the operators $T_{21}(\buc_{\st})$ onto the Bethe vector. Then due to
\eqref{A-T21} we obtain a sum over partitions of the set $\{\buc_{\st},\bub\}$. The terms of this sum  are proportional to the
products of the functions $r_1(\bet_{\so})$, where $\bet_{\so}\subset \{\buc_{\st},\bub\}$ and $\#\bet_{\so}=a-n$. Hence, the
parameters $\bet_{\so}$ are of the $u$-type.

Next, we act with the operators $\bT_{31}(\buc_{\so})$ onto obtained Bethe vectors via \eqref{A-T31}. We get new partitions of the set
$\bigl\{\{\buc,\bub\}\setminus\bet_{\so}\bigr\}$ and new products of functions $r_1$, say, $r_1(\bet_{\so'})$. Obviously,
$\bet_{\so'}\subset \bigl\{\{\buc,\bub\}\setminus\bet_{\so}\bigr\}$ and $\#\bet_{\so'}=n$. Thus, the total number of the
functions $r_1$ is equal to $a$, and all their arguments are of the $u$-type.

Finally, we should act with the product of the operators $\bT_{32}(\bvc_{\st})$. But due to \eqref{A-T32} this action does not
produce new functions $r_1$. Thus, we have proved that $\bw_{\qo}\subset \{\buc,\bub\}$ and $\#\bw_{\qo}=a$.

Similarly, one can prove that $\bw_{\qt}\subset \{\bvc,\bvb\}$ and $\#\bw_{\qt}=b$. However, for this one should take representation
\eqref{dPhi-expl1} for the dual Bethe vector. Then all the functions $r_3$ will be produced under the successive actions of the operators
$\bT_{32}(\bvc_{\st})$ and $\bT_{31}(\bvc_{\so})$. Repeating the considerations above we prove that all the parameters
$\bw_{\qt}$ are of the $v$-type and their total number is equal to $b$.\qed

Due to proposition~\ref{uv-type} one can recast \eqref{Sab-genform} in the form
\be{Sab-genform1}
S_{a,b}(\buc;\bvc|\bub;\bvb)=\sum r_1(\bet_{\so})r_3(\bxi_{\so})W_{\text{part}}(\bet_{\so};\bet_{\st}|\bxi_{\so};\bxi_{\st}).
\ee
Here $\bet=\{\buc,\bub\}$ and $\bxi=\{\bvc,\bvb\}$. The sum is taken over partitions
$\bet\Rightarrow\{\bet_{\so},\bet_{\st}\}$ and $\bxi\Rightarrow\{\bxi_{\so},\bxi_{\st}\}$, such that
$\#\bet_{\so}=a$ and $\#\bxi_{\so}=b$. Setting in \eqref{Sab-genform1}
\be{newsets}
\begin{array}{lll}
\bet_{\so}=\{\bub_{\so},\;\buc_{\st}\}, &\qquad \bet_{\st}=\{\bub_{\st},\;\buc_{\so}\},&\qquad \#\bub_{\so}=\#\buc_{\so}=k,\quad k=1,\dots,a;\\
\bxi_{\so}=\{\bvb_{\so},\;\bvc_{\st}\} , &\qquad  \bxi_{\st}=\{\bvb_{\st},\;\bvc_{\so}\},&\qquad \#\bvb_{\so}=\#\bvc_{\so}=n,\quad n=1,\dots,b ,
\end{array}
\ee
we arrive at a representation
\be{Sab-genform2}
S_{a,b}(\buc;\bvc|\bub;\bvb)=\sum \frac{r_1(\buc_{\st})r_1(\bub_{\so})r_3(\bvc_{\st})r_3(\bvb_{\so})}
{f(\bvc,\buc)f(\bvb,\bub)}
W_{\text{part}}\begin{pmatrix}\buc_{\st},\bub_{\st},&\buc_{{\so}},\bub_{{\so}}\\
\bvc_{\so},\bvb_{\so},&\bvc_{\st},\bvb_{\st}\end{pmatrix}.
\ee
Here the sum runs over all the partitions $\buc\Rightarrow\{\buc_{\so},\buc_{\st}\}$,
$\bub\Rightarrow\{\bub_{\so},\bub_{\st}\}$,  $\bvc\Rightarrow\{\bvc_{\so},\bvc_{\st}\}$ and $\bvb\Rightarrow\{\bvb_{\so},\bvb_{\st}\}$
with $\#\buc_{\so}=\#\bub_{\so}$ and $\#\bvc_{\so}=\#\bvb_{\so}$.
The functions $W_{\text{part}}$ are rational coefficients, which depend on the partitions but do not
depend on the functions $r_1$ and $r_3$. We also have extracted explicitly the product $f(\bvc,\buc)^{-1}f(\bvb,\bub)^{-1}$ that plays
the role of a normalization factor.
\begin{Def}
We call the highest coefficient $Z_{a,b}(\buc;\bub|\bvc;\bvb)$ the function $W_{\text{\rm part}}$  that corresponds to the extreme partitions
$\buc_{\so}=\bub_{\so}=\emptyset$ and $\bvc_{\st}=\bvb_{\st}=\emptyset$:
\be{def:Zl}
W_{\text{\rm part}}\begin{pmatrix}\buc,\bub,&\emptyset,\emptyset\\
\bvc,\bvb,&\emptyset,\emptyset\end{pmatrix}=Z_{a,b}(\buc;\bub|\bvc;\bvb).
\ee
In other words, $Z_{a,b}(\buc;\bub|\bvc;\bvb)$ is the coefficient of the product $r_1(\buc)r_3(\bvb)$.
\end{Def}

One also can define a conjugated highest coefficient corresponding to the extreme partition $\buc_{\st}=\bub_{\st}=\emptyset$ and $\bvc_{\so}=\bvb_{\so}=\emptyset$, that  is the coefficient of the product $r_1(\bub)r_3(\bvc)$. However, due to \eqref{SP-CB} it is clear that
\be{def:Zr}
W_{\text{part}}\begin{pmatrix}\emptyset,\emptyset,&\buc,\bub,\\
\emptyset,\emptyset,&\bvc,\bvb\end{pmatrix}=Z_{a,b}(\bub;\buc|\bvb;\bvc).
\ee
We will show that all other coefficients $W_{\text{part}}$ are equal to bilinear combinations of the highest coefficient and its conjugated.

\begin{prop}\label{W-2lin}
For a fixed partition with $\#\buc_{\so}=\#\bub_{\so}=k$ and $\#\bvc_{\so}=\#\bvb_{\so}=n$, (where $k=0,\dots,a$ and
$n=0,\dots,b$), the coefficient $W_{\text{\rm part}}$  has the form

 \begin{multline}\label{W-Reshet}
W_{\text{\rm part}}\begin{pmatrix}\buc_{\st},\bub_{\st},&\buc_{\so},\bub_{\so}\\
\bvc_{\so},\bvb_{\so},&\bvc_{\st},\bvb_{\st}\end{pmatrix}=f(\bub_{\st},\bub_{\so}) f(\buc_{\so},\buc_{\st})g(\bvb_{\st},\bvb_{\so})g(\bvc_{\so},\bvc_{\st})
f(\bvc_{\so},\buc_{\so}) f(\bvb_{\st},\bub_{\st})\num
\times
Z_{a-k,n}(\buc_{\st};\bub_{\st}|\bvc_{\so};\bvb_{\so}) \;Z_{k,b-n}(\bub_{\so};\buc_{\so}|\bvb_{\st};\bvc_{\st})\;.
 \end{multline}
\end{prop}

The main goal of this paper is to find an explicit formula for the highest coefficient $Z_{a,b}$ and to prove the representation \eqref{W-Reshet} for the coefficient $W_{\text{\rm part}}$.

Comparing \eqref{W-Reshet} with the analogous formula for the $\mathfrak{gl}(3)$-based models \cite{Res86} one can see that they are
very similar. It is enough to replace the product $g(\bvb_{\st},\bvb_{\so})g(\bvc_{\so},\bvc_{\st})$ in \eqref{W-Reshet} with the product
$f(\bvb_{\so},\bvb_{\st})f(\bvc_{\st},\bvc_{\so})$ and we obtain the formula of the paper \cite{Res86}. One should remember, however, that
the highest coefficients also have different representations in the models described by $\mathfrak{gl}(2|1)$ and $\mathfrak{gl}(3)$ algebras.
In particular, we will see that in the case under consideration the highest coefficient $Z_{a,b}$ admits a single determinant representation,
while in the $\mathfrak{gl}(3)$ case such a determinant formula is not known.

\section{Successive actions\label{S-SA}}

In the previous section we have described how the scalar product depends on the functions $r_k$. Our goal now is to find
explicitly the rational coefficients $W_{\text{part}}$. For this we calculate successive action of the operators
$T_{ij}$ with $i>j$ onto a generic Bethe vector. This calculation is quite similar to the one of the work \cite{HutLPRS16a},
where we computed the multiple actions of the monodromy matrix entries.

\subsection{Successive action of $\bT_{31}(\bx)T_{21}(\by)$}

We start with the successive action of the products $\bT_{31}(\bx)T_{21}(\by)$.
Let $\#\bar x=n$ and  $\#\bar y=a-n$ where $n=0,1,\dots,\min(a,b)$. Define
\be{def-G}
G_{n,a}(\bx,\by)=\frac{\bT_{31}(\bx)T_{21}(\by)}{\lambda_{2}(\bx)\lambda_{2}(\by)}\mathbb{B}_{a,b}(\bu;\bv).
\ee
Using successively \eqref{A-T21} and \eqref{A-T31} we obtain
\begin{multline}\label{A-T3121}
G_{n,a}(\bx,\by)=(-1)^{\frac{n(n+1)}2}h(\bv,\by)h(\by,\by)\sum r_1(\bet_{\so})
\frac{f(\bet_{\st},\bet_{\so})f(\bet_{\st},\bet_{\sth})f(\bet_{\sth},\bet_{\so})g(\bxi_{\st},\bxi_{\so})}
{h(\bxi_{\so},\by)f(\bxi_{\st},\bet_{\so})}\\
\times K_{a-n}(\by|\bet_{\st}+c)K_{a-n}(\bet_{\so}|\bxi_{\so}+c)
h(\bxi_{\st},\bx)h(\bx,\bx)r_3(\bxi_{\rm i})r_1(\bet_{\rm ii})
\frac{g(\bxi_{\rm ii},\bxi_{\rm i})g(\bxi_{\rm iii},\bxi_{\rm ii})g(\bxi_{\rm iii},\bxi_{\rm i})}
{h(\bet_{\rm i},\bx)h(\bxi_{\rm i},\bet_{\rm i})h(\bxi_{\rm ii},\bx)}\\
\times \frac{f(\bet_{\rm i},\bet_{\rm ii})
h(\bet_{\rm i},\bet_{\rm i})}{f(\bxi_{\rm i},\bet_{\rm ii})f(\bxi_{\rm iii},\bet_{\rm ii})}
\;K_n(\bet_{\rm ii}|\bxi_{\rm ii}+c)\;\mathbb{B}_{0,b-n}(\emptyset;\bxi_{\rm iii}).
\end{multline}
The sum is organized as follows. First the sets $\{\by,\bu\}$ and $\{\by,\bv\}$ are divided respectively into subsets
$\{\bet_{\so},\bet_{\st},\bet_{\sth}\}$ and $\{\bxi_{\so},\bxi_{\st}\}$
with the restriction $\#\bxi_{\so}=\#\bet_{\so}=\#\bet_{\st}=a-n$. Then the sets $\{\bx,\bet_{\sth}\}$ and $\{\bx,\bxi_{\st}\}$
are divided respectively into subsets
$\{\bet_{\rm i},\bet_{\rm ii}\}$ and $\{\bxi_{\rm i},\bxi_{\rm ii},\bxi_{\rm iii}\}$ with the restriction
$\#\bxi_{\rm i}=\#\bxi_{\rm ii}=\#\bet_{\rm i}=\#\bet_{\rm ii}=n$.

It is convenient to introduce the sets $\bet=\{\by,\bx,\bu\}$ and $\bxi=\{\by,\bx,\bv\}$. Then we can understand the sum in
\eqref{A-T3121} as the sum over partitions  $\bet\Rightarrow\{\bet_{\so},\bet_{\st},\bet_{\rm i},\bet_{\rm ii}\}$ and
$\bxi\Rightarrow\{\bxi_{\so},\bxi_{\rm i},\bxi_{\rm ii},\bxi_{\rm iii}\}$ with the restrictions mentioned above and an additional
constrain $\bx\cap\{\bet_{\so},\bet_{\st},\bxi_{\so}\} =\emptyset$. Hereby $\bet_{\sth}= \{\bet_{\rm i},\bet_{\rm ii}\}\setminus \bx$
and $\bxi_{\st} = \{\bxi_{\rm i},\bxi_{\rm ii},\bxi_{\rm iii}\}\setminus \bx$. Then we have
\be{project-1}
f(\bet_{\st},\bet_{\sth})f(\bet_{\sth},\bet_{\so})=\frac{f(\bet_{\st},\bet_{\rm i})f(\bet_{\st},\bet_{\rm ii})
f(\bet_{\rm i},\bet_{\so})f(\bet_{\rm ii},\bet_{\so})}{f(\bet_{\st},\bx)f(\bx,\bet_{\so})},
\ee
and
\be{project-2}
\frac{g(\bxi_{\st},\bxi_{\so})}{f(\bxi_{\st},\bet_{\so})}=
\frac{g(\bxi_{\rm i},\bxi_{\so})g(\bxi_{\rm ii},\bxi_{\so})g(\bxi_{\rm iii},\bxi_{\so})f(\bx,\bet_{\so})}
{g(\bx,\bxi_{\so})f(\bxi_{\rm i},\bet_{\so})f(\bxi_{\rm ii},\bet_{\so})f(\bxi_{\rm iii},\bet_{\so})}.
\ee
Observe that the restrictions $\bx\cap\bet_{\st} =\emptyset$ and $\bx\cap\bxi_{\so} =\emptyset$ hold automatically
due to the presence of the product $f(\bet_{\st},\bx)$ in the denominator of \eqref{project-1} and the product $g(\bx,\bxi_{\so})$ in the
denominator of \eqref{project-2}. Indeed, $1/f(\bet_{\st},\bx)=0$ as soon as $\bx\cap\bet_{\st} \ne \emptyset$ and $1/g(\bx,\bxi_{\so})=0$ as soon
as $\bx\cap\bxi_{\so} \ne\emptyset$. Actually, one can easily see that the condition $\bx\cap\bet_{\so} =\emptyset$ also holds, although
the product $f(\bx,\bet_{\so})$ in the denominator of \eqref{project-1} is compensated by the same product in the numerator of
\eqref{project-2}. Indeed, we have seen that $\bx\cap\bxi_{\so} =\emptyset$, that is to say, $\bx\subset\{\bxi_{\rm i},\bxi_{\rm ii},\bxi_{\rm iii}\}$.
But in this case $\bx\cap\bet_{\so} =\emptyset$ due to the products of the $f$-functions in the denominator of \eqref{project-2}.

Thus, we can recast \eqref{A-T3121} as follows:
\begin{multline}\label{A-T3121-1}
G_{n,a}(\bx,\by)=(-1)^{\frac{n(n+1)}2}\frac{h(\bxi,\by)h(\bxi,\bx)}{h(\bx,\by)}\sum r_1(\bet_{\so})r_1(\bet_{\rm ii})r_3(\bxi_{\rm i})\num
\times
\frac{f(\bet_{\st},\bet_{\so}) f(\bet_{\st},\bet_{\rm i}) f(\bet_{\st},\bet_{\rm ii})
f(\bet_{\rm i},\bet_{\so}) f(\bet_{\rm ii},\bet_{\so}) f(\bet_{\rm i},\bet_{\rm ii})h(\bet_{\rm i},\bet_{\rm i})}
{f(\bet_{\st},\bx)f(\bxi_{\rm i},\bet_{\so}) f(\bxi_{\rm ii},\bet_{\so})f(\bxi_{\rm iii},\bet_{\so})
f(\bxi_{\rm i},\bet_{\rm ii}) f(\bxi_{\rm iii},\bet_{\rm ii}) }\num
\times
\frac{g(\bxi_{\rm i},\bxi_{\so})g(\bxi_{\rm ii},\bxi_{\so})g(\bxi_{\rm iii},\bxi_{\so})
g(\bxi_{\rm ii},\bxi_{\rm i})g(\bxi_{\rm iii},\bxi_{\rm ii})g(\bxi_{\rm iii},\bxi_{\rm i})}
{h(\bxi_{\so},\by)h(\bxi_{\so},\bx)h(\bet_{\rm i},\bx)h(\bxi_{\rm i},\bet_{\rm i})h(\bxi_{\rm ii},\bx)
g(\bx,\bxi_{\so})}\num
\times
K_{a-n}(\by|\bet_{\st}+c) K_{a-n}(\bet_{\so}|\bxi_{\so}+c)\;K_n(\bet_{\rm ii}|\bxi_{\rm ii}+c)\;\mathbb{B}_{0,b-n}(\emptyset;\bxi_{\rm iii}).
\end{multline}
Here we have also used
\be{h-h}
h(\bx,\bx)h(\bxi_{\st},\bx)=\frac{h(\bxi,\bx)}{h(\bxi_{\so},\bx)}.
\ee
In \eqref{A-T3121-1} the sum is taken over partitions  $\bet\Rightarrow\{\bet_{\so},\bet_{\st},\bet_{\rm i},\bet_{\rm ii}\}$ and
$\bxi\Rightarrow\{\bxi_{\so},\bxi_{\rm i},\bxi_{\rm ii},\bxi_{\rm iii}\}$. The restriction are imposed on the cardinalities of
the subsets only.

One can reduce the number of subsets in \eqref{A-T3121-1}. Let $\bet_{0}=\{\bet_{\so},\bet_{\rm ii}\}$. Then \eqref{A-T3121-1} takes the form
\begin{multline}\label{A-T3121-1a}
G_{n,a}(\bx,\by)=(-1)^{\frac{n(n+1)}2}\frac{h(\bxi,\by)h(\bxi,\bx)}{h(\bx,\by)}\sum r_1(\bet_{0})r_3(\bxi_{\rm i})
\frac{ f(\bet_{\st},\bet_{0})f(\bet_{\rm i},\bet_{0})}{f(\bxi_{\rm iii},\bet_{0})
f(\bxi_{\rm i},\bet_{0})}\;K_{a-n}(\by|\bet_{\st}+c)\num
\times
\frac{f(\bet_{\st},\bet_{\rm i})h(\bet_{\rm i},\bet_{\rm i})   g(\bxi_{\rm i},\bxi_{\so})g(\bxi_{\rm ii},\bxi_{\so})g(\bxi_{\rm iii},\bxi_{\so})
g(\bxi_{\rm ii},\bxi_{\rm i})g(\bxi_{\rm iii},\bxi_{\rm ii})g(\bxi_{\rm iii},\bxi_{\rm i})}
{f(\bet_{\st},\bx)   h(\bxi_{\so},\by)h(\bxi_{\so},\bx)h(\bet_{\rm i},\bx)h(\bxi_{\rm i},\bet_{\rm i})h(\bxi_{\rm ii},\bx)
g(\bx,\bxi_{\so})}\;\mathbb{B}_{0,b-n}(\emptyset;\bxi_{\rm iii})\num
\times
\frac{f(\bet_{\rm ii},\bet_{\so})}{f(\bxi_{\rm ii},\bet_{\so})}
K_{a-n}(\bet_{\so}|\bxi_{\so}+c)\;K_n(\bet_{\rm ii}|\bxi_{\rm ii}+c).
\end{multline}
We see that the sum over partitions $\bet_{0}\Rightarrow\{\bet_{\so},\bet_{\rm ii}\}$ involves the  terms in the last line
only. This sum can be computed via lemma~\ref{main-ident}. Using \eqref{Sym-Part-old1} we find
\begin{multline}\label{Appl-ML}
\sum_{\bet_{0}\Rightarrow\{\bet_{\so},\bet_{\rm ii}\}}
K_{a-n}(\bet_{\so}|\bxi_{\so}+c)\;K_n(\bet_{\rm ii}|\bxi_{\rm ii}+c)\frac{f(\bet_{\rm ii},\bet_{\so})}{f(\bxi_{\rm ii},\bet_{\so})}\\
=\frac{(-1)^n}{f(\bxi_{\rm ii},\bet_{0})}
\sum_{\bet_{0}\Rightarrow\{\bet_{\so},\bet_{\rm ii}\}}
K_{a-n}(\bet_{\so}|\bxi_{\so}+c)\;K_n(\bxi_{\rm ii}|\bet_{\rm ii})f(\bet_{\rm ii},\bet_{\so})
=\frac{(-1)^a K_a(\{\bxi_{\so},\bxi_{\rm ii}\}|\bet_{0})}{f(\bxi_{\rm ii},\bet_{0})f(\bxi_{\so},\bet_{0})}.
\end{multline}
Thus, \eqref{A-T3121-1a} takes the form
\begin{multline}\label{A-T3121-2}
G_{n,a}(\bx,\by)=(-1)^{a+\frac{n(n+1)}2}\frac{h(\bxi,\by)h(\bxi,\bx)}{h(\bx,\by)}\sum r_1(\bet_{0})r_3(\bxi_{\rm i})
K_{a-n}(\by|\bet_{\st}+c)\num
\times
\frac{f(\bet_{\st},\bet_{0}) f(\bet_{\st},\bet_{\rm i})
f(\bet_{\rm i},\bet_{0})h(\bet_{\rm i},\bet_{\rm i})
g(\bxi_{\rm i},\bxi_{\so})g(\bxi_{\rm ii},\bxi_{\so})g(\bxi_{\rm iii},\bxi_{\so})
g(\bxi_{\rm ii},\bxi_{\rm i})g(\bxi_{\rm iii},\bxi_{\rm ii})g(\bxi_{\rm iii},\bxi_{\rm i})}
{f(\bet_{\st},\bx) f(\bxi_{\rm ii},\bet_{0})f(\bxi_{\so},\bet_{0}) f(\bxi_{\rm iii},\bet_{0})
f(\bxi_{\rm i},\bet_{0})
h(\bxi_{\so},\by)h(\bxi_{\so},\bx)h(\bet_{\rm i},\bx)h(\bxi_{\rm i},\bet_{\rm i}) h(\bxi_{\rm ii},\bx)
g(\bx,\bxi_{\so}) }\num
\times K_a(\{\bxi_{\so},\bxi_{\rm ii}\}|\bet_{0})\;\mathbb{B}_{0,b-n}(\emptyset;\bxi_{\rm iii}).
\end{multline}

Now we define $\bxi_{0}=\{\bxi_{\so},\bxi_{\rm ii}\}$. Then \eqref{A-T3121-2} can be written as
\begin{multline}\label{A-T3121-2a}
G_{n,a}(\bx,\by)=(-1)^{a+\frac{n(n-1)}2}\frac{h(\bxi,\by)h(\bxi,\bx)}{h(\bx,\by)}\sum r_1(\bet_{0})r_3(\bxi_{\rm i})
K_{a-n}(\by|\bet_{\st}+c)K_a(\bxi_{0}|\bet_{0})\num
\times
\frac{f(\bet_{\st},\bet_{0}) f(\bet_{\st},\bet_{\rm i})
f(\bet_{\rm i},\bet_{0})h(\bet_{\rm i},\bet_{\rm i})
g(\bxi_{\rm i},\bxi_{0})g(\bxi_{\rm iii},\bxi_{0})g(\bxi_{\rm iii},\bxi_{\rm i})}
{f(\bet_{\st},\bx) f(\bxi_{0},\bet_{0}) f(\bxi_{\rm iii},\bet_{0})
f(\bxi_{\rm i},\bet_{0})h(\bet_{\rm i},\bx)h(\bxi_{\rm i},\bet_{\rm i})h(\bxi_{0},\bx)}
\;\mathbb{B}_{0,b-n}(\emptyset;\bxi_{\rm iii}) \num
\times\frac{g(\bxi_{\rm ii},\bxi_{\so})}{ h(\bxi_{\so},\by)g(\bx,\bxi_{\so})}.
\end{multline}
The sum over partitions $\bxi_{0}\Rightarrow\{\bxi_{\so},\bxi_{\rm ii}\}$ involves
the  terms in the last line only. It can be computed via \eqref{Sym-Part-new1}:
\begin{multline}\label{anal-ML}
\sum \frac{g(\bxi_{\rm ii},\bxi_{\so})}{g(\bx,\bxi_{\so})h(\bxi_{\so},\by)}=\frac{(-1)^{n(a-n)}}{g(\bxi_{0},\bx)}
\sum \frac{g(\bxi_{\rm ii},\bxi_{\so})g(\bxi_{\rm ii},\bx)}{h(\bxi_{\so},\by)}\\
=\frac{(-1)^{n(a-n)}}{g(\bxi_{0},\bx)}
\sum g(\bxi_{\rm ii},\bxi_{\so})g(\bxi_{\rm ii},\bx)g(\bxi_{\so},\by-c)=\frac{h(\bx,\by)}{h(\bxi_{0},\by)}.
\end{multline}
Thus, we arrive at
\begin{multline}\label{A-T3121-3}
G_{n,a}(\bx,\by)=(-1)^{a+\frac{n(n-1)}2}h(\bxi,\by)h(\bxi,\bx)\sum \frac{r_1(\bet_{0})r_3(\bxi_{\rm i})}
{h(\bxi_{0},\bx)h(\bxi_{0},\by)}
K_{a-n}(\by|\bet_{\st}+c) K_a(\bxi_{0}|\bet_{0})\num
\times
\frac{f(\bet_{\st},\bet_{0}) f(\bet_{\st},\bet_{\rm i})
f(\bet_{\rm i},\bet_{0})h(\bet_{\rm i},\bet_{\rm i})   g(\bxi_{\rm i},\bxi_{0}) g(\bxi_{\rm iii},\bxi_{0})
g(\bxi_{\rm iii},\bxi_{\rm i})}
{f(\bet_{\st},\bx) f(\bxi_{0},\bet_{0}) f(\bxi_{\rm iii},\bet_{0})
f(\bxi_{\rm i},\bet_{0})    h(\bet_{\rm i},\bx)h(\bxi_{\rm i},\bet_{\rm i})  }
\;\mathbb{B}_{0,b-n}(\emptyset;\bxi_{\rm iii}).
\end{multline}
Finally, after a relabeling of the subsets
$\bet_{0}\to \bet_{\so}$, $\bet_{\rm i}\to \bet_{\st}$, $\bet_{\st}\to \bet_{\sth}$,
$\bxi_{\rm i}\to\bxi_{\so}$, $\bxi_{0}\to\bxi_{\st}$, $\bxi_{\rm iii}\to\bxi_{\sth}$,
we recast \eqref{A-T3121-3} as follows:
\begin{multline}\label{A-T3121-4}
G_{n,a}(\bx,\by)=(-1)^{a+\frac{n(n-1)}2}h(\bxi,\by)h(\bxi,\bx)\sum \frac{r_1(\bet_{\so})r_3(\bxi_{\so})}
{h(\bxi_{\st},\bx)h(\bxi_{\st},\by)}
K_{a-n}(\by|\bet_{\sth}+c) K_a(\bxi_{\st}|\bet_{\so})\num
\times
\frac{f(\bet_{\sth},\bet_{\so}) f(\bet_{\sth},\bet_{\st})
f(\bet_{\st},\bet_{\so})h(\bet_{\st},\bet_{\st})    g(\bxi_{\so},\bxi_{\st}) g(\bxi_{\sth},\bxi_{\st})
g(\bxi_{\sth},\bxi_{\so})}
{f(\bet_{\sth},\bx) f(\bxi,\bet_{\so}) h(\bet_{\st},\bx)h(\bxi_{\so},\bet_{\st})  }
\;\mathbb{B}_{0,b-n}(\emptyset;\bxi_{\sth}).
\end{multline}
We recall that the cardinalities of the subsets are
\be{card-new0}
\begin{aligned}
\#\bet_{\so}=a,\qquad & \#\bet_{\st}=n,\qquad & \#\bet_{\sth}=a-n,\\
\#\bxi_{\so}=n,\qquad & \#\bxi_{\st}=a,\qquad & \#\bxi_{\sth}=b-n.
\end{aligned}
\ee
\begin{rmk}\label{REM-lim}
Strictly speaking, the sets $\bar\eta$ and $\bar\xi$ in  equation \eqref{A-T3121-4} should be understood as
\be{eps}
\begin{array}{l}
\bar\eta=\{\bx+\epsilon_1,\by+\epsilon_1, \bu+\epsilon_1\},\\
\bar\xi=\{\bx+\epsilon_2,\by+\epsilon_2, \bv+\epsilon_2\},
\end{array}
\qquad\text{at}\quad \epsilon_{k}\to 0,\qquad k=1,2.
\ee
The point is that individual
factors in \eqref{A-T3121-4}  may have singularities, if we set $\epsilon_k=0$. For instance, if $\bxi_{\st}\cap\bet_{\so}\ne\emptyset$,
then the DWPF $K_a(\bxi_{\st}|\bet_{\so})$ is singular. However, these poles are compensated by the product $f(\bxi,\bar\eta_{\so})^{-1}$.
Therefore, for appropriate evaluating the limit we should have $\epsilon_k\ne 0$. In order to lighten the formulas we do
not write these auxiliary parameters $\epsilon_k$ explicitly, but  one has to keep them in mind  when doing the calculations.
\end{rmk}

\subsection{Successive action of $\bT_{32}(\bz)\bT_{31}(\bx)T_{21}(\by)$}

Let now  $\#\bar z=b-n$. Then we define
\be{ewa-act}
\frac{\bT_{32}(\bar z)\, \bT_{31}(\bar x)\, T_{21}(\bar y)}{\lambda_2(\bar z)\, \lambda_2(\bar x)\,\lambda_2(\bar y)}
\mathbb{B}_{a,b}(\bu,\bv)=\frac{\bT_{32}(\bar z)}{\lambda_2(\bar z)}G_{n,a}(\bx,\by)=H_{n,a,b}(\bx,\by,\bz)|0\rangle.
\ee
In order to act with $\bT_{32}(\bar z)$ onto $G_{n,a}(\bx,\by)$ we should use \eqref{A-T32b2}. Let us denote the union
$\{\bx,\by\}$ as $\buc$ (as it will be in the case of the scalar product). Then we obtain
\begin{multline}\label{A-T32T3121-1}
H_{n,a,b}(\bx,\by,\bz)=(-1)^{a+\frac{n(n-1)}2+\frac{(b-n)(b-n-1)}2}h(\bv,\buc)h(\buc,\buc)
\sum \frac{r_1(\bet_{\so})r_3(\bxi_{\so})r_3(\bxi_{\rm i})}{h(\bxi_{\st},\buc)}
\num
\times K_{a-n}(\by|\bet_{\sth}+c) K_a(\bxi_{\st}|\bet_{\so})
\frac{f(\bet_{\sth},\bet_{\so}) f(\bet_{\sth},\bet_{\st})
f(\bet_{\st},\bet_{\so})h(\bet_{\st},\bet_{\st})}
{f(\bet_{\sth},\bx) f(\bxi,\bet_{\so}) }\\
\times \frac{ g(\bxi_{\so},\bxi_{\st}) g(\bxi_{\sth},\bxi_{\st})
g(\bxi_{\sth},\bxi_{\so})}
{h(\bet_{\st},\bx)h(\bxi_{\so},\bet_{\st})  }g(\bxi_{\rm ii},\bxi_{\rm i}).
\end{multline}
Here the partitions of the set $\bet$ remain the same as in \eqref{A-T3121-4}. The partitions of the remaining variables
are organized  as follows. We first have the partitions of the set $\{\buc,\bv\}=\bxi\Rightarrow\{\bxi_{\so},\bxi_{\st},\bxi_{\sth}\}$.
Then we combine $\{\bz,\bxi_{\sth}\}$ and obtain additional partitions $\{\bz,\bxi_{\sth}\}\Rightarrow\{\bxi_{\rm i},\bxi_{\rm ii}\}$ with
the restriction $\#\bxi_{\rm i}=\#\bxi_{\rm ii}=b-n$.

We should substitute $\bxi_{\sth}=\{\bxi_{\rm i},\bxi_{\rm ii}\}\setminus \bz$ into \eqref{A-T32T3121-1}. Then, using
\be{subs-use}
\frac{g(\bxi_{\sth},\bxi_{\st})g(\bxi_{\sth},\bxi_{\so})}{f(\bxi,\bet_{\so})}=
\frac{g(\bxi_{\rm i},\bxi_{\st})g(\bxi_{\rm ii},\bxi_{\st})g(\bxi_{\rm i},\bxi_{\so})g(\bxi_{\rm ii},\bxi_{\so})}
{f(\bv,\bet_{\so})f(\buc,\bet_{\so})g(\bz,\bxi_{\so})g(\bz,\bxi_{\st})},
\ee
we arrive at
\begin{multline}\label{A-T32T3121-2}
H_{n,a,b}(\bx,\by,\bz)=(-1)^{a+\frac{n(n-1)}2+\frac{(b-n)(b-n-1)}2}h(\bv,\buc)h(\buc,\buc)
\sum \frac{r_1(\bet_{\so})r_3(\bxi_{\so})r_3(\bxi_{\rm i})  }{h(\bxi_{\st},\buc)}
\num
\times K_{a-n}(\by|\bet_{\sth}+c) K_a(\bxi_{\st}|\bet_{\so})
\frac{f(\bet_{\sth},\bet_{\so}) f(\bet_{\sth},\bet_{\st})
f(\bet_{\st},\bet_{\so})h(\bet_{\st},\bet_{\st})f(\bz,\bet_{\so})}
{f(\bet_{\sth},\bx) f(\bxi,\bet_{\so}) }\\
\times \frac{g(\bxi_{\so},\bxi_{\st})g(\bxi_{\rm i},\bxi_{\st}) g(\bxi_{\rm ii},\bxi_{\st})
g(\bxi_{\rm i},\bxi_{\so})   g(\bxi_{\rm ii},\bxi_{\so})g(\bxi_{\rm ii},\bxi_{\rm i})}
{h(\bet_{\st},\bx)  h(\bxi_{\so},\bet_{\st})g(\bz,\bxi_{\so})g(\bz,\bxi_{\st}) }.
\end{multline}
Here we have denoted by $\bxi$ the union $\{\bz,\buc,\bv\}$. This set is divided into four subsets
$\bxi\Rightarrow\{\bxi_{\rm i},\bxi_{\rm ii},\bxi_{\so},\bxi_{\st}\}$ with the cardinalities
$\#\bxi_{\rm i}=\#\bxi_{\rm ii}=b-n$, $\#\bxi_{\so}=n$, and $\#\bxi_{\st}=a$.

Let $\bxi_{0}=\{\bxi_{\rm i},\bxi_{\so}\}$.  Then
\begin{multline}\label{A-T32T3121-2a}
H_{n,a,b}(\bx,\by,\bz)=(-1)^{a+n(b+1)+\frac{b(b-1)}2}h(\bv,\buc)h(\buc,\buc)
\sum \frac{r_1(\bet_{\so}) r_3(\bxi_{0})  }{h(\bxi_{\st},\buc)}K_{a-n}(\by|\bet_{\sth}+c)
\num
\times  K_a(\bxi_{\st}|\bet_{\so})
\frac{f(\bet_{\sth},\bet_{\so}) f(\bet_{\sth},\bet_{\st})
f(\bet_{\st},\bet_{\so})h(\bet_{\st},\bet_{\st})f(\bz,\bet_{\so})
g(\bxi_{0},\bxi_{\st}) g(\bxi_{\rm ii},\bxi_{\st})
 g(\bxi_{\rm ii},\bxi_{0}) }
{f(\bet_{\sth},\bx) f(\bxi,\bet_{\so}) h(\bet_{\st},\bx)  g(\bz,\bxi_{\st}) }\\
\times
\frac{g(\bxi_{\rm i},\bxi_{\so}) }{ h(\bxi_{\so},\bet_{\st})g(\bz,\bxi_{\so})}.
\end{multline}
The sum over partitions $\bxi_{0}\Rightarrow\{\bxi_{\so},\bxi_{\rm i}\}$ involves
the  terms in the last line only. It can be computed via \eqref{Sym-Part-new1}:
\be{sum-OM}
\sum\frac{g(\bxi_{\rm i},\bxi_{\so}) } {h(\bxi_{\so},\bet_{\st})g(\bz,\bxi_{\so})}=
\frac{(-1)^{b-n}}{g(\bz,\bxi_{0})}\sum\frac{g(\bxi_{\rm i},\bxi_{\so})g(\bxi_{\rm i},\bz) } {h(\bxi_{\so},\bet_{\st})}
=\frac{h(\bz,\bet_{\st})}{h(\bxi_{0},\bet_{\st})}.
\ee
Substituting this into \eqref{A-T32T3121-2a} we find
\begin{multline}\label{A-T32T3121-3}
H_{n,a,b}(\bx,\by,\bz)=(-1)^{a+n(b+1)+\frac{b(b-1)}2}h(\bv,\buc)h(\buc,\buc)
\sum \frac{r_1(\bet_{\so}) r_3(\bxi_{0}) }{h(\bxi_{\st},\buc)}
\num
\times
\frac{f(\bet_{\sth},\bet_{\so}) f(\bet_{\sth},\bet_{\st})
f(\bet_{\st},\bet_{\so})h(\bet_{\st},\bet_{\st})f(\bz,\bet_{\so})g(\bxi_{0},\bxi_{\st})
g(\bxi_{\rm ii},\bxi_{\st})g(\bxi_{\rm ii},\bxi_{0})h(\bz,\bet_{\st})}
{f(\bet_{\sth},\bx) f(\bxi,\bet_{\so})h(\bet_{\st},\bx) h(\bxi_{0},\bet_{\st})g(\bz,\bxi_{\st}) }\num
\times
K_{a-n}(\by|\bet_{\sth}+c) K_a(\bxi_{\st}|\bet_{\so}).
\end{multline}
Finally, relabeling $\bxi_{0}\to \bxi_{\so}$ and $\bxi_{\rm ii}\to \bxi_{\sth}$ we arrive at
\begin{multline}\label{H-1}
H_{n,a,b}(\bx,\by,\bz)=(-1)^{a+n(b+1)+\frac{b^2-b}2}\frac{h(\bxi,\buc)}{h(\bar z,\buc)}\sum r_1(\bet_{\so})r_3(\bxi_{\so})
\,K_a(\bxi_{\st}|\bet_{\so}) K_{a-n}(\bar y|\bet_{\sth}+c)\\
\times\frac{f(\bet_{0},\bet_{\so})h(\bet_{0},\bet_{\st})g(\bet_{\sth},\bet_{\st})f(\bar z,\bet_{\so})h(\bar z,\bet_{\st})
g(\bxi_{\so},\bxi_{\st}) g(\bxi_{\sth},\bxi_{\so}) g(\bxi_{\sth},\bxi_{\st})   }
{ h(\bxi_{\so},\bet_{\st}) f(\bxi,\bet_{\so}) h(\bet_{\st},\bar x)  f(\bet_{\sth},\bar x)g(\bar z ,\bxi_{\st})h(\bxi_{\st},\buc)    }.
\end{multline}
Recall that in this formula $\bet=\{\bar x,\bar y,\bu\}$, $\bxi=\{\bar z, \bar x,\bar y,\bv\}$, and we denote $\buc=\{\bx,\by\}$.
The sum is taken over partitions
$\bet\Rightarrow\{\bet_{\so},\bet_{\st},\bet_{\sth}\}$ and $\bxi\Rightarrow\{\bxi_{\so},\bxi_{\st},\bxi_{\sth}\}$.
Hereby $\bet_{0}=\{\bet_{\st},\bet_{\sth}\}$. The cardinalities of the subsets are
\be{card}
\begin{aligned}
&\#\bet_{\so}=a,\qquad &\#\bet_{\st}=n,\qquad &\#\bet_{\sth}=a-n,\\
&\#\bxi_{\so}=b,\qquad &\#\bxi_{\st}=a,\qquad &\#\bxi_{\sth}=b-n.
\end{aligned}
\ee

\section{Highest coefficients\label{S-HC}}

Equation \eqref{H-1} allows us to obtain an explicit representation for the scalar product of Bethe vectors. Using
\eqref{dPhi-expl2} we find
\be{SP-first}
S_{a,b}=\frac{ (-1)^{\frac{b^2-b}2} }{f(\bvc,\buc)}
\sum g(\bvc_{\so},\buc_{\so}) f(\buc_{\st},\buc_{\so})f(\bvc_{\so},\buc_{\st}) g(\bvc_{\st},\bvc_{\so})h(\buc_{\so},\buc_{\so})
\;H_{n,a,b}(\buc_{\so},\buc_{\st},\bvc_{\st}).
\ee
The sum is taken over partitions $\buc\Rightarrow\{\buc_{\so},\buc_{\st}\}$ and $\bvc\Rightarrow\{\bvc_{\so},\bvc_{\st}\}$,
where $\#\bvc_{\so}=\#\buc_{\so}=n$, and $n=0,1,\dots,\min(a,b)$. The function $H_{n,a,b}(\buc_{\so},\buc_{\st},\bvc_{\st})$ itself
is given as a sum over partitions described in \eqref{H-1}. Namely, the union $\{\buc,\bub\}$ is divided into three subsets and
the union $\{\bvc_{\st},\buc,\bvb\}$ also is divided into three subsets. Although  the resulting formula is explicit, it is inconvenient for later use.
Therefore, we will try to simplify it. To do this, we introduce a new function.

\begin{Def}
Let $\bx$, $\by$, $\bt$, $\bs$, and $\bar\beta$ be five sets of generic complex numbers with cardinalities
$\#\bar x=n$, $\#\bar y=m$, and $\#\bar \beta=n+m$. The cardinalities of the sets $\bar t$ and $\bar s$ are not fixed.
Define a  function
\be{def-J}
J_{n,m}(\bar x;\bar y|\bar t;\bar s|\bar\beta)=
\Delta_{n+m}(\bar\beta)\Delta'_n(\bar x)\Delta'_m(\bar y)\;
\det_{n+m}\mathcal{J}_{jk},
\ee
where
\be{matelJ}
\begin{aligned}
&\mathcal{J}_{jk}=\frac{g(\beta_j,x_k)}{h(\beta_j,x_k)},&\qquad k=1,\dots,n;\\
&\mathcal{J}_{j,k+n}=g(\beta_j,y_k)\frac{h(\beta_j,\bar t)}{h(\beta_j,\bar s)},&\qquad k=1,\dots,m;
\end{aligned}
\qquad j=1,\dots,n+m.
\ee
\end{Def}

Developing the determinant in \eqref{def-J} with respect to the first $n$ columns (see Appendix~\ref{A-SF} for more details) we obtain a presentation of $J_{n,m}$ as a sum
over partitions of the set $\bbet$:
\be{J-sum}
J_{n,m}(\bar x;\bar y|\bar t;\bar s|\bar\beta)=\sum K_n(\bbet_{\so}|\bx)\frac{g(\bbet_{\st},\bbet_{\so})g(\bbet_{\st},\by)h(\bbet_{\st},\bt)}
{h(\bbet_{\so},\bx)h(\bbet_{\st},\bs)}.
\ee
Here the sum is taken over partitions $\bbet\Rightarrow\{\bbet_{\so},\bbet_{\st}\}$, such that $\#\bbet_{\so}=n$ and $\#\bbet_{\st}=m$.

\subsection{First representation for the highest coefficient}

Let us find the highest coefficient $Z_{a,b}(\bub;\buc|\bvb;\bvc)$. We recall that up to the normalization factor
$\bigl(f(\bvc,\buc)f(\bvb,\bub)\bigr)^{-1}$ it is the rational coefficient of the product $r_1(\bub)r_3(\bvc)$
(see \eqref{def:Zr}, \eqref{W-Reshet}).

Obviously, for this we should set $\bet_{\so}=\bub$ and $\bxi_{\so}=\bvc$ in \eqref{H-1}. However,
$\bxi_{\so}\subset\{\bvc_{\st},\buc,\bvb\}$. Hence, one can have $\bxi_{\so}=\bvc$ if and only if $\bvc_{\st}=\bvc$, and thus,
$\bvc_{\so}=\emptyset$. But $\#\bvc_{\so}=\#\buc_{\so}=n$ in \eqref{SP-first}, therefore $\buc_{\so}=\emptyset$ and
$n=0$. Thus, \eqref{SP-first} takes the form
\be{HC-rep1}
\frac{r_1(\bub)r_3(\bvc)\;Z_{a,b}(\bub;\buc|\bvb;\bvc)}{f(\bvc,\buc)f(\bvb,\bub)}=\frac{ (-1)^{\frac{b^2-b}2} }{f(\bvc,\buc)}
\;H_{0,a,b}(\emptyset,\buc,\bvc)\Bigr|_{\bet_{\so}=\bub;~\bxi_{\so}=\bvc}.
\ee
Substituting the conditions $\bet_{\so}=\bub$ and $\bxi_{\so}=\bvc$  into \eqref{H-1} we should take into account that
$\#\bet_{\st}=n=0$ (see \eqref{card}). Hence, $\bet_{\st}=\emptyset$, which implies $\bet_{\sth}=\bet_{0}=\buc$. Thus,
substituting these subsets into \eqref{H-1} we find
\begin{multline}\label{Z-pc1}
r_1(\bub)r_3(\bvc)\;Z_{a,b}(\bub;\buc|\bvb;\bvc)=(-1)^{a}h(\bvb,\buc)h(\buc,\buc) r_1(\bub)r_3(\bvc)
\\
\times K_{a}(\buc|\buc+c)\sum K_a(\bxi_{\st}|\bub)
\frac{ g(\bxi_{\sth},\bvc) g(\bxi_{\sth},\bxi_{\st})   }
{  h(\bxi_{\st},\buc)    },
\end{multline}
where the sum is taken over partitions $\{\buc,\bvb\}=\bxi\Rightarrow\{\bxi_{\st},\bxi_{\sth}\}$ with $\#\bxi_{\st}=a$ and
$\#\bxi_{\sth}=b$. Due to \eqref{K-K} we conclude that $K_{a}(\buc|\buc+c)=(-1)^a$, and we arrive at
\begin{equation}\label{Z-pc2}
Z_{a,b}(\bub;\buc|\bvb;\bvc)=h(\bvb,\buc)h(\buc,\buc)\sum K_a(\bxi_{\st}|\bub)
\frac{ g(\bxi_{\sth},\bvc) g(\bxi_{\sth},\bxi_{\st})   }
{  h(\bxi_{\st},\buc)    }.
\end{equation}
Finally, using $\{\buc,\bvb\}=\bxi$ we recast \eqref{Z-pc2} as follows:
\begin{multline}\label{Z-pc3}
Z_{a,b}(\bub;\buc|\bvb;\bvc)=\sum K_a(\bxi_{\st}|\bub)
 g(\bxi_{\sth},\bvc) g(\bxi_{\sth},\bxi_{\st}) h(\bxi_{\sth},\buc)\\
=h(\bvb,\bub)h(\buc,\bub)\sum K_a(\bxi_{\st}|\bub)
\frac{ g(\bxi_{\sth},\bvc) g(\bxi_{\sth},\bxi_{\st}) h(\bxi_{\sth},\buc)   }
{  h(\bxi_{\st},\bub)h(\bxi_{\sth},\bub)    }.
\end{multline}
Comparing \eqref{Z-pc3} and \eqref{J-sum} we conclude that
\be{Z-J}
Z_{a,b}(\bub;\buc|\bvb;\bvc)=h(\bvb,\bub)h(\buc,\bub)\;J_{a,b}(\bub;\bvc|\buc;\bub|\{\bvb,\buc\}).
\ee
Thus, we have obtained an explicit  representation for the highest coefficient $Z_{a,b}(\bub;\buc|\bvb;\bvc)$
in terms of the determinant of the $(a+b)\times(a+b)$ matrix $\mathcal{J}_{jk}$ \eqref{matelJ}.

\subsection{Second highest coefficient}

In order to obtain the second highest coefficient $Z_{a,b}(\buc;\bub|\bvc;\bvb)$ it is enough to make the
replacements $\buc\leftrightarrow\bub$ and $\bvc\leftrightarrow\bvb$ in \eqref{Z-J}. On the other hand, this coefficient should
arise if we  set $\bet_{\so}=\buc$ and $\bxi_{\so}=\bvb$ in \eqref{H-1}. However, if we do so, then we do not obtain
\eqref{Z-J} with the replacements mentioned above. Instead, we obtain much more sophisticated formula involving many sums over partitions. This `break of symmetry'
occurs because we use a specific representation \eqref{dPhi-expl2} for the dual Bethe vector. If we would use equation \eqref{dPhi-expl1}
for $\mathbb{C}_{a,b}(\bu;\bv)$, then we would have an analog of \eqref{Z-J} for $Z_{a,b}(\buc;\bub|\bvc;\bvb)$, however, we would
have a more complex formula for $Z_{a,b}(\bub;\buc|\bvb;\bvc)$.

A `complex' formula for the highest coefficient provides
us with a very non-trivial identity for   $Z_{a,b}(\buc;\bub|\bvc;\bvb)$, that will be used later. In order to obtain this
identity we first make several additional summations in \eqref{SP-first}. Let us rewrite this equation explicitly
\begin{multline}\label{SP-expl1}
S_{a,b}=\frac{  h(\bvb,\buc)h(\buc,\buc)} {f(\bvc,\buc)}
\sum (-1)^{a+n(b+1)} g(\bvc_{\so},\buc) g(\buc_{\st},\buc_{\so})h(\bvc_{\so},\buc_{\st}) g(\bvc_{\st},\bvc_{\so})h(\buc,\buc_{\so})\\
\times r_1(\bet_{\so})r_3(\bxi_{\so})
\,K_a(\bxi_{\st}|\bet_{\so}) K_{a-n}(\buc_{\st}|\bet_{\sth}+c)\\
\times\frac{f(\bet_{0},\bet_{\so})h(\bet_{0},\bet_{\st})g(\bet_{\sth},\bet_{\st})f(\bvc_{\st},\bet_{\so})h(\bvc_{\st},\bet_{\st})
g(\bxi_{\so},\bxi_{\st}) g(\bxi_{\sth},\bxi_{\so}) g(\bxi_{\sth},\bxi_{\st})   }
{ h(\bxi_{\so},\bet_{\st}) f(\bxi,\bet_{\so}) h(\bet_{\st},\buc_{\so})  f(\bet_{\sth},\buc_{\so})g(\bvc_{\st} ,\bxi_{\st})h(\bxi_{\st},\buc)    }.
\end{multline}
The sum over partitions into subsets $\bxi_{\st}$ and $\bxi_{\sth}$, as well as the sum over partitions $\buc\Rightarrow\{\buc_{\so},\buc_{\st}\}$
can be computed in terms of
the function $J$ \eqref{def-J}. Let $\bxi_{0}=\{\bxi_{\st},\bxi_{\sth}\}$. Then
\begin{multline}\label{sum-xi}
\sum_{\bxi_{0}\Rightarrow\{\bxi_{\st},\bxi_{\sth}\}}
\frac{ K_a(\bxi_{\st}|\bet_{\so})g(\bxi_{\so},\bxi_{\st}) g(\bxi_{\sth},\bxi_{\so}) g(\bxi_{\sth},\bxi_{\st})}
{g(\bvc_{\st} ,\bxi_{\st})h(\bxi_{\st},\buc)}\\
=\frac{(-1)^{ab+n+b}g(\bxi_{0},\bxi_{\so})h(\bxi_{0},\bet_{\so})}
{g(\bvc_{\st},\bxi_{0}) h(\bxi_{0},\buc)}J_{a,b-n}(\bet_{\so};\bvc_{\st}| \buc; \bet_{\so}|\bxi_{0}).
\end{multline}
Similarly, one can verify that
\begin{multline}\label{sum-u}
\sum_{\buc\Rightarrow\{\buc_{\so},\buc_{\st}\}}\frac{ K_{a-n}(\buc_{\st}|\bet_{\sth}+c)
 g(\buc_{\st},\buc_{\so})h(\bvc_{\so},\buc_{\st}) h(\buc,\buc_{\so})}
 {h(\bet_{\st},\buc_{\so})f(\bet_{\sth},\buc_{\so})}\\
=(-1)^{a+n+an}\frac{h(\bvc_{\so},\buc)}{g(\bet_{\sth},\buc)}J_{a-n,n}(\bet_{\sth};\bvc_{\so}| \buc+c; \bet_{0}+c|\buc-c).
\end{multline}
Substituting these results into \eqref{SP-expl1} we find
\begin{multline}\label{SP-expl2}
S_{a,b}=\frac{  h(\bvb,\buc)h(\buc,\buc)} {f(\bvc,\buc)}
\sum (-1)^{b+ab+n(a+b+1)} r_1(\bet_{\so})r_3(\bxi_{\so})J_{a,b-n}(\bet_{\so};\bvc_{\st}| \buc; \bet_{\so}|\bxi_{0})\\
\times J_{a-n,n}(\bet_{\sth};\bvc_{\so}| \buc+c; \bet_{0}+c|\buc-c)f(\bet_{0},\bet_{\so})h(\bet_{0},\bet_{\st})g(\bet_{\sth},\bet_{\st})
\\
\times\frac{ h(\bvc_{\st},\bet_{\st})f(\bvc_{\so},\buc)  g(\bvc_{\st},\bvc_{\so})
g(\bxi_{0},\bxi_{\so})h(\bxi_{0},\bet_{\so})  }
{g(\bet_{\sth},\buc) f(\buc,\bet_{\so}) f(\bvb,\bet_{\so}) h(\bxi_{\so},\bet_{\st}) g(\bvc_{\st},\bxi_{0})h(\bxi_{0},\buc)    }.
\end{multline}
The sum is taken over partitions:
\\
\begin{tabular}{l}
(1) $\bvc\Rightarrow\{\bvc_{\so},\bvc_{\st}\}$ with $\#\bvc_{\so}=n$ and $\#\bvc_{\st}=b-n$;\\
(2) $\{\buc,\bvb,\bvc_{\st}\}=\bxi\Rightarrow\{\bxi_{\so},\bxi_{0}\}$ with $\#\bxi_{\so}=b$ and $\#\bxi_{0}=b+a-n$;\\
(3) $\{\buc,\bub\}=\bet\Rightarrow\{\bet_{\so},\bet_{0}\}$ and $\bet_{0}\Rightarrow\{\bet_{\st},\bet_{\sth}\}$ with
$\#\bet_{\so}=a$, $\#\bet_{\st}=n$, and $\#\bet_{\sth}=a-n$.
\end{tabular}\\
In all these partitions $n=0,1,\dots,\min(a,b)$.

Due to proposition~\ref{uv-type} the function $r_3$ depends on the variables of the $v$-type only. Hence, $\bxi_{\so}\cap
\buc=\emptyset$, that is $\buc\subset\bxi_{0}$. Therefore, we can set $\bxi_{0}=\{\buc,\bxi_{\st}\}$. Substituting this
into \eqref{SP-expl2} we obtain
\begin{multline}\label{SP-expl3}
S_{a,b}=\frac{  h(\bvb,\buc)} {f(\bvc,\buc)}
\sum (-1)^{b+n(a+b+1)} r_1(\bet_{\so})r_3(\bxi_{\so})J_{a,b-n}(\bet_{\so};\bvc_{\st}| \buc; \bet_{\so}|\{\bxi_{\st},\buc\})\\
\times J_{a-n,n}(\bet_{\sth};\bvc_{\so}| \buc+c; \bet_{0}+c|\buc-c)f(\bet_{0},\bet_{\so})h(\bet_{0},\bet_{\st})g(\bet_{\sth},\bet_{\st})
\\
\times\frac{ h(\bvc_{\st},\bet_{\st})f(\bvc_{\so},\buc)  g(\bvc_{\st},\bvc_{\so})
g(\bxi_{\st},\bxi_{\so})h(\bxi_{\st},\bet_{\so})g(\bxi_{\so},\buc)  }
{g(\bet_{\sth},\buc) g(\buc,\bet_{\so}) f(\bvb,\bet_{\so}) h(\bxi_{\so},\bet_{\st}) g(\bvc_{\st},\bxi_{\st})h(\bxi_{\st},\buc)g(\bvc_{\st},\buc)  }.
\end{multline}
In this formula $\bxi=\{\bvc_{\st},\bvb\}$, $\#\bxi_{\so}=b$ and $\#\bxi_{\st}=b-n$. All other subsets are the same as in \eqref{SP-expl2}.

Now everything is ready to formulate the second representation for the highest coefficient  $ Z_{a,b}(\buc;\bub|\bvc;\bvb)$.
For this we set $\bet_{\so}=\buc$ and $\bxi_{\so}=\bvb$. Then
automatically
$\bet_0=\bub$, $\bxi_{\st}=\bvc_{\st}$ and we also can set $\bet_{\st}=\bub_{\so}$, $\bet_{\sth}=\bub_{\st}$.
Substituting this into \eqref{SP-expl3} and keeping in mind remark \ref{REM-lim} we obtain
\begin{multline}\label{SP-3pc1}
Z_{a,b}(\buc;\bub|\bvc;\bvb)=f(\bvb,\bub)f(\bub,\buc)\sum (-1)^{n(a+b+1)+b}
  g(\bvc_{\st},\bvc_{\so})f(\bvc_{\so},\buc) \\
\times  J_{a-n,n}(\bub_{\st};\bvc_{\so}|\buc+c;\bub+c|\buc-c)\frac{h(\bub,\bub_{\so})g(\bub_{\st},\bub_{\so})h(\bvc_{\st},\bub_{\so})
g(\bvc_{\st},\bvb) h(\bvc_{\st},\buc)    }
{g(\bub_{\st},\buc)h(\bvb,\bub_{\so})g(\bvc_{\st} ,\buc)h(\bvc_{\st},\buc)    }
\, \num
\times \lim_{\substack{\bet_{\so}\to\buc\\ \bxi_{\st}\to\bvc_{\st}}}
\frac{J_{a,b-n}(\bet_{\so};\bvc_{\st}| \buc; \bet_{\so}|\{\buc,\bxi_{\st}\})}{g(\buc,\bet_{\so})g(\bvc_{\st} ,\bxi_{\st})}.
\end{multline}
Using \eqref{limW-1} and \eqref{limW-2} we find
\be{lim}
\lim_{\substack{\bet_{\so}\to\buc\\ \bxi_{\st}\to\bvc_{\st}}}
\frac{J_{a,b-n}(\bet_{\so};\bvc_{\st}| \buc; \bet_{\so}|\{\buc,\bxi_{\st}\})}{g(\buc,\bet_{\so})g(\bvc_{\st} ,\bxi_{\st})}
=(-1)^{b+n}g(\bvc_{\st} ,\buc),
\ee
and we thus arrive at
\begin{multline}\label{Ident-1-main}
Z_{a,b}(\buc;\bub|\bvc;\bvb)= f(\bub,\buc)f(\bvb,\bub)\sum (-1)^{n(a+b)}
J_{a-n,n}(\bub_{\st};\bvc_{\so}|\buc+c;\bub+c|\buc-c)\\
\times h(\bub_{\so},\bub_{\so})f(\bub_{\st},\bub_{\so})\;\frac{f(\bvc_{\so},\buc) h(\bvc_{\st},\bub_{\so})
g(\bvc_{\st},\bvc_{\so})g(\bvc_{\st},\bvb)}
{h(\bvb,\bub_{\so})g(\bub_{\st},\buc)}.
\end{multline}
Here the sum is taken over partitions $\bub\Rightarrow\{\bub_{\st},\bub_{\so}\}$  and $\bvc\Rightarrow\{\bvc_{\so},\bvc_{\st}\}$
with $\#\bvc_{\so}=n$ and $\#\bub_{\st}=a-n$.

This is the second representation for the highest coefficient discussed above. It will play the key role below, therefore we formulate
it as a proposition.

\begin{prop}\label{non-tr-prop}
For arbitrary sets of complex numbers $\bt$, $\bx$, $\bs$, and $\by$ with cardinalities
$\#\bt=\#\bx=a$ and $\#\bs=\#\by=b$ the following identity holds:
\begin{multline}\label{Ident-0-main}
Z_{a,b}(\bt;\bx|\bs;\by)= f(\bx,\bt)f(\by,\bx)\sum (-1)^{n_1(a+b)}
J_{\ell_2,n_1}(\bx_{\st};\bs_{\so}|\bt+c;\bx+c|\bt-c)\\
\times h(\bx_{\so},\bx_{\so})f(\bx_{\st},\bx_{\so})\;\frac{f(\bs_{\so},\bt) h(\bs_{\st},\bx_{\so})
g(\bs_{\st},\bs_{\so})g(\bs_{\st},\by)}
{h(\by,\bx_{\so})g(\bx_{\st},\bt)}.
\end{multline}
Here $\ell_2=\#\bx_{\st}$, $n_1=\#\bs_{\so}$. The sum is taken over partitions
\be{part-0}
\bx\Rightarrow\{\bx_{\so},\bx_{\st}\},  \qquad \bs\Rightarrow\{\bs_{\so},\bs_{\st}\},
\ee
with a restriction $\#\bs_{\so}=\#\bx_{\so}$ (which is equivalent to $\ell_2+n_1=a$).
\end{prop}

{\sl Proof}. Setting in \eqref{Ident-1-main} $\buc=\bt$, $\bub=\bx$, $\bvc=\bs$, and $\bvb=\by$ we obtain
\eqref{Ident-0-main}.

\subsection{General formula for the scalar product}

Now we turn back to equation \eqref{SP-expl3}. To proceed further we should specify all the subsets. Let
$\bvc=\{\bvc_{\qo},\bvc_{\qt},\bvc_{\qth}\}$ and $\bvb=\{\bvb_{\qo},\bvc_{\qt}\}$. We set
\be{part-v}
\begin{aligned}
& \bvc_{\so}=\bvc_{\qth},\qquad & \bvc_{\st}=\{\bvc_{\qo},\bvc_{\qt}\}, \\
& \bvb=\{\bvb_{\qo},\bvb_{\qt}\},\qquad &{} \\
&\bxi_{\so}=\{\bvc_{\qo},\bvb_{\qt}\},\qquad &\bxi_{\st}=\{\bvc_{\qt},\bvb_{\qo}\},
\end{aligned}
\mb{ with }
\begin{cases}
 \#\bvc_s=n_s,\\
 \#\bvb_s=m_s,
\end{cases}
\qquad s=\qo, \qt,\qth.
\ee
 It is easy to see that the
following conditions for the cardinalities hold:
\be{restr-v}
n_{\qth}=n,\quad n_{\qo}+n_{\qt}=b-n, \quad m_{\qo}+m_{\qt}=b, \quad m_{\qo}=n_{\qo}.
\ee
Let also
\be{part-u}
\begin{aligned}
& \buc=\{\buc_{\qo},\buc_0\},\qquad & \buc_{0}=\{\buc_{\qt},\buc_{\qth}\},\\
& \bub=\{\bub_{\qo},\bub_0\},\qquad &  \bub_{\qo}=\{\bub_{\qt},\bub_{\qth}\}, \\
& \bet_{\so}=\{\buc_{\qo},\bub_0\},\qquad & \bet_{0}=\{\buc_0,\bub_{\qo}\},%
 %\bet_{\st}=\{\buc_{\qt},\bub_{\qth}\},\qquad & \bet_{\sth}=\{\buc_{\qth},\bub_{\qt}\}.\qquad \\
\end{aligned}
\mb{ with }
\begin{cases} \#\buc_p=k_p, \\ \#\bub_p=\ell_p, \end{cases}
\qquad p=0,\qo, \qt,\qth.
\ee
We have the following conditions for the cardinalities:
\be{restr-u}
k_{\qo}+k_0=\ell_{\qo}+\ell_0=a,\quad k_{\qo}=\ell_{\qo}, \quad k_0=\ell_0.%, \quad k_2+\ell_3=n,\quad k_3+\ell_2=a-n.
\ee
Observe that we do not fix a distribution of the parameters $\buc$ and $\bub$ among the subsets $\bet_{\st}$
and $\bet_{\sth}$. It is important, however, that $\#\bet_{\st}=n=n_{\qth}$
and $\#\bet_{\sth}=a-n$.

Using \eqref{limW-1} and \eqref{limW-2} we obtain
\begin{multline}\label{red-W}
\frac{J_{a,b-n}(\bet_{\so};\bvc_{\st}| \buc; \bet_{\so}|\{\buc,\bxi_{\st}\})}
{g(\buc,\bet_{\so})g(\bvc_{\st} ,\bxi_{\st}) }=(-1)^{b-n}
\frac{g(\bvc_{\qt},\buc)g(\bvb_{\qo},\buc_{\qo}) h(\bvc_{\qt},\buc_0)}
{g(\buc_0,\bub_0)g(\bvb_{\qo} ,\bvc_{\qo}) h(\bvc_{\qt},\bub_0) }\\
\times J_{\ell_0,n_{\qo}}(\bub_0;\bvc_{\qo}| \buc_0; \bub_0|\{\buc_0,\bvb_{\qo}\}).
\end{multline}
Due to \eqref{Z-J} this function reduces to the highest coefficient
\be{W-Z1}
J_{\ell_0,n_{\qo}}(\bub_0;\bvc_{\qo}| \buc_0; \bub_0|\{\buc_0,\bvb_{\qo}\})=
\frac{Z_{\ell_0,n_{\qo}}(\bub_0,\buc_0|\bvb_{\qo},\bvc_{\qo})}
{h(\buc_0,\bub_0)h(\bvb_{\qo},\bub_0)}.
\ee

Now we substitute \eqref{red-W}, \eqref{W-Z1} into \eqref{SP-expl3}. We also write explicitly the products $g(\bvc_{\st},\bvc_{\so})$,
$g(\bxi_{\st},\bxi_{\so})$, $f(\bet_0,\bet_{\so})$, and combine $\{\bvc_{\qt},\bvc_{\qth}\}=\bvc_0$. Then we have
\begin{multline}\label{SP-4}
S_{a,b}=\frac{1}{ f(\bvc,\buc)}\sum
r_1(\buc_{\qo})r_1(\bub_0)r_3(\bvc_{\qo})r_3(\bvb_{\qt}) \;Z_{\ell_0,n_{\qo}}(\bub_0,\buc_0|\bvb_{\qo},\bvc_{\qo})
 \frac{f(\bub_{\qo},\buc_{\qo})f(\bvb_{\qt},\buc)}
{f(\bvb,\bub_{0})f(\bvb_{\qt},\buc_{\qo})}\\
\times    f(\buc_0,\buc_{\qo})f(\bub_{\qo},\bub_0)g(\bvc_0,\bvc_{\qo})g(\bvb_{\qo},\bvb_{\qt})\quad
\Bigl\{  J_{a-n,n}(\bet_{\sth};\bvc_{\qth}|\buc+c;\bet_{0}+c|\buc-c) \num
   \times  (-1)^{n(a+b-n_{\qo})} h(\bet_{\st},\bet_{\st})f(\bet_{\sth},\bet_{\st})
\frac{f(\bvc_{\qth},\buc)g(\bvc_{\qt},\bvc_{\qth})g(\bvc_{\qt},\bvb_{\qt})h(\bvc_{\qt},\bet_{\st})  }
{ g(\bet_{\sth},\buc)h(\bvb_{\qt},\bet_{\st})   }\Bigr\}.
\end{multline}
Here the sum is organized as follows. First, we  have partitions
\be{Afin-part}
\begin{aligned}
&\buc\Rightarrow\{\buc_{\qo},\buc_0\},\qquad &\bub\Rightarrow\{\bub_{\qo},\bub_0\},\qquad &\#\buc_0=\#\bub_0=\ell_0,\\
&\bvc\Rightarrow\{\bvc_{\qo},\bvc_0\},\qquad &\bvb\Rightarrow\{\bvb_{\qo},\bvb_{\qt}\}, \qquad &\#\bvc_{\qo}=\#\bvb_{\qo}=n_{\qo}.
\end{aligned}
\ee
After this we have two additional partitions: the set $\bvc_0$ is divided into subsets $\bvc_{\qt}$ and $\bvc_{\qth}$;
the union of the subsets $\bet_0=\{\buc_0,\bub_{\qo}\}$ is divided into subsets $\bet_{\st}$ and $\bet_{\sth}$ (see the terms in braces in \eqref{SP-4}).
Hereby we have one restriction for the cardinalities $\#\bvc_{\qth}=\#\bet_{\st}=n=n_{\qth}$. Let us write separately this additional sum over partitions
 in braces of \eqref{SP-4}  
\begin{multline}\label{AddSum}
 \mathcal{F}(\buc;\bet_0;\bvc_0,\bvb_{\qt})= \sum_{\substack{\bet_0 \Rightarrow \{\bet_{\st},\bet_{\sth}\}\\
 \bvc_0 \Rightarrow \{\bvc_{\qt},\bvc_{\qth}\}  } }
 J_{a-n,n}(\bet_{\sth};\bvc_{\qth}|\buc+c;\bet_{0}+c|\buc-c) \num
   \times  (-1)^{n(a+b-n_{\qo})} h(\bet_{\st},\bet_{\st})f(\bet_{\sth},\bet_{\st})
\frac{f(\bvc_{\qth},\buc)g(\bvc_{\qt},\bvc_{\qth})g(\bvc_{\qt},\bvb_{\qt})h(\bvc_{\qt},\bet_{\st})  }
{ g(\bet_{\sth},\buc)h(\bvb_{\qt},\bet_{\st})   }.
\end{multline}

Comparing \eqref{AddSum} with \eqref{Ident-0-main} one can see that they coincide after appropriate identification of
the subsets and their cardinalities. Namely, \eqref{AddSum} turns into \eqref{Ident-0-main} under the
replacements $b-n_{\qo} \to b$, $\buc\to\bt$, $\bet_0\to\bx$, $\bvc_0\to\bs$, $\bvb_{\qt}\to\by$. Thus, due to Proposition~\ref{non-tr-prop}
we obtain
\begin{equation}\label{sum-nontr}
\mathcal{F}(\buc;\bet_0;\bvc_0,\bvb_{\qt})=\frac{Z_{a,b-n_{\qo}}(\buc;\bet_0|\bvc_0,\bvb_{\qt})}
{f(\bet_0,\buc)f(\bvb_{\qt},\bet_0)}.
\end{equation}
Thus, substituting this into \eqref{SP-4} we arrive at
\begin{multline}\label{SP-5}
S_{a,b}=\frac{1}{ f(\bvc,\buc)}\sum
r_1(\buc_{\qo})r_1(\bub_0)r_3(\bvc_{\qo})r_3(\bvb_{\qt}) \;Z_{\ell_0,n_{\qo}}(\bub_0,\buc_0|\bvb_{\qo},\bvc_{\qo})
 \frac{f(\bub_{\qo},\buc_{\qo})f(\bvb_{\qt},\buc)}
{f(\bvb,\bub_{0})f(\bvb_{\qt},\buc_{\qo})}\num
\times    f(\buc_0,\buc_{\qo})f(\bub_{\qo},\bub_0)g(\bvc_0,\bvc_{\qo})g(\bvb_{\qo},\bvb_{\qt})\,
\frac{Z_{a,b-n_{\qo}}(\buc;\bet_0|\bvc_0,\bvb_{\qt})}
{f(\bet_0,\buc)f(\bvb_{\qt},\bet_0)}.
\end{multline}
It remains to simplify the ratio $Z_{a,b-n_{\qo}}(\buc;\bet_0|\bvc_0,\bvb_{\qt})/f(\bet_0,\buc)$. It can be done  via
 \eqref{Z-J}, \eqref{limW-1}:
\be{red-Z2}
\frac{Z_{a,b-n_{\qo}}(\buc;\bet_0|\bvc_0,\bvb_{\qt})}
{f(\bet_0,\buc)f(\bvb_{\qt},\bet_0)}=\frac{f(\bvc_0,\buc_0)Z_{a-\ell_0,b-n_{\qo}}(\buc_{\qo};\bub_{\qo}|\bvc_0,\bvb_{\qt})}
{f(\bub_{\qo},\buc_{\qo})f(\bvb_{\qt},\bub_{\qo})f(\bvb_{\qt},\buc_0)}.
\ee
Substituting this into \eqref{SP-5} we obtain
\begin{multline}\label{Sab-exp}
S_{a,b}=
\sum  r_1(\buc_{\qo})r_1(\bub_0)r_3(\bvc_{\qo})r_3(\bvb_{\qt})f(\buc_0,\buc_{\qo})f(\bub_{\qo},\bub_0)g(\bvc_0,\bvc_{\qo})g(\bvb_{\qo},\bvb_{\qt})\\
\times \frac{f(\bvc_0,\buc_0)f(\bvb_{\qo},\bub_{\qo})}{ f(\bvc,\buc)f(\bvb,\bub)}\;
 Z_{\ell_0,n_{\qo}}(\bub_0,\buc_0|\bvb_{\qo},\bvc_{\qo})\; Z_{a-\ell_0,b-n_{\qo}}(\buc_{\qo},\bub_{\qo}|\bvc_0,\bvb_{\qt}).
\end{multline}
It is easy to see that after appropriate relabeling the subsets we arrive at
\begin{multline}\label{Sab-fin}
S_{a,b}=
\sum  r_1(\buc_{\st})r_1(\bub_{\so})r_3(\bvc_{\st})r_3(\bvb_{\so})f(\buc_{\so},\buc_{\st})f(\bub_{\st},\bub_{\so})g(\bvc_{\so},\bvc_{\st})g(\bvb_{\st},\bvb_{\so})\\
\times \frac{f(\bvc_{\so},\buc_{\so})f(\bvb_{\st},\bub_{\st})}{ f(\bvc,\buc)f(\bvb,\bub)}\;
  Z_{a-k,n}(\buc_{\st},\bub_{\st}|\bvc_{\so},\bvb_{\so})\;Z_{k,b-n}(\bub_{\so},\buc_{\so}|\bvb_{\st},\bvc_{\st}),
\end{multline}
where $k=\#\buc_{\so}=\#\bub_{\so}$ and $n=\#\bvc_{\so}=\#\bvb_{\so}$.
Comparing this result with \eqref{Sab-genform2} and \eqref{W-Reshet} we see that proposition~\ref{W-2lin} is proved.

\section{Scalar product in the $\mathfrak{gl}(1|1)$ sector\label{S-SP11}}

Consider a particular case of the subalgebra $\mathfrak{gl}(1|1)$,  generated by the operators $T_{23}(u)$, $T_{22}(u)$,
$T_{33}(u)$ and $T_{32}(u)$. In this case one should set $\buc=\bub=\emptyset$ in
the formulas for the scalar product. Then the highest coefficient simplifies as
\be{Z-det11}
Z_{0,b}(\emptyset;\emptyset|\bs;\by)=\Delta_{b}(\bs)\Delta'_b(\bar y)\;
\det_{b}g(s_j,y_k)=g(\bs,\by),
\ee
where we used an explicit representation for Cauchy determinant
\be{Cauchy-0}
\det_m g(u_j,v_k)=\frac{g(\bu,\bv)}{\Delta'_m(\bu)\Delta_m(\bv)}.
\ee
The scalar product \eqref{Sab-fin} takes the form
\begin{equation}\label{Sa0-fin}
S_{0,b}=
\sum  r_3(\bvc_{\st})r_3(\bvb_{\so})g(\bvc_{\so},\bvc_{\st})g(\bvb_{\st},\bvb_{\so})
g(\bvc_{\so},\bvb_{\so})g(\bvb_{\st},\bvc_{\st}),
\end{equation}
where the sum is taken over partitions $\bvc\Rightarrow\{\bvc_{\so},\bvc_{\st}\}$ and $\bvb\Rightarrow\{\bvb_{\so},\bvb_{\st}\}$
such that $\#\bvc_{\so}=\bvb_{\so}$. It is easy to see that this sum reduces to a single determinant
\begin{equation}\label{Sa0-det}
S_{0,b}=\Delta'_b(\bvc)\Delta_b(\bvb)\det_b\Bigl[g(\vc_j,\vb_k)\bigl(r_3(\vb_k)-r_3(\vc_j)\bigr)\Bigr].
\end{equation}
Indeed, developing the determinant in \eqref{Sa0-det} via Laplace formula and using \eqref{Cauchy-0}, \eqref{Del-sig} we
obtain the sum \eqref{Sa0-fin}.

Thus, the scalar product of Bethe vectors in  $\mathfrak{gl}(1|1)$ integrable models admits a determinant representation
without any restriction on the Bethe parameters. This is not surprising, as these models are equivalent to free fermions
\cite{FoeK93,GohM98}.

\section{Different representations for the highest coefficient\label{S-DRHC}}

If $\bvc=\bvb=\emptyset$, then formula \eqref{Sab-fin} describes the scalar product in the $\mathfrak{gl}(2)$-based models.
In this case the scalar product admits a determinant representation, if one of the Bethe vectors is an eigenvector of the transfer matrix.
One  expects that in the general $\mathfrak{gl}(2|1)$ case the sum over partitions in \eqref{Sab-fin} also can be reduced to a single
determinant for some particular cases of Bethe vectors. To make this reduction
one should have different representations for the highest coefficient
$ Z_{a,b}(\bar t;\bar x|\bar s;\bar y)$. In this section we give several formulas for $Z_{a,b}$ in terms of sums over partitions and multiple
contour integrals.

We have already obtained an expression for $Z_{a,b}$ as the determinant of an $(a+b)\times(a+b)$ matrix
\be{Z-detJ}
Z_{a,b}(\bt;\bx|\bs;\by)=h(\bw,\bt)\Delta_{a+b}(\bw)\Delta'_a(\bar t)\Delta'_b(\bar y)\;
\det_{a+b}\mathcal{J}_{jk},
\ee
where $\bw=\{\bx,\bs\}$ and the matrix $\mathcal{J}_{jk}$ is defined in \eqref{matelJ}:
\be{matelJ1}
\begin{aligned}
&\mathcal{J}_{jk}=\frac{g(w_j,t_k)}{h(w_j,t_k)},&\qquad k=1,\dots,a;\\
&\mathcal{J}_{j,k+a}=g(w_j,y_k)\frac{h(w_j,\bar x)}{h(w_j,\bar t)},&\qquad k=1,\dots,b;
\end{aligned}
\qquad j=1,\dots,a+b.
\ee
Developing the determinant with respect to the $a$ first columns we
obtain
 \be{RHC-IHC}
 Z_{a,b}(\bt;\bx|\bs;\by)=\sum K_{a}(\bar w_{\so}|\bar t) h(\bar w_{\st},\bar x) g(\bar w_{\st},\bar y) g(\bar w_{\st},\bar w_{\so}).
 \ee
The sum is taken over partitions $\{\bx,\bs\}=\bw\Rightarrow\{\bar w_{\so},\bar w_{\st}\}$ with $\#\bar w_{\so}=a$ and $\#\bar w_{\st}=b$.

Let us give several alternative representations for the highest coefficient.
\begin{itemize}
\item As a sum over partitions of  $\bar t$ and $\bar y$:
 \be{Al-RHC-IHC}
 Z_{a,b}(\bar t;\bar x|\bar s;\bar y)=f(\bar s,\bar t)f(\bar y,\bar x)
   \sum g(\bar\eta_{\so},\bar\eta_{\st}) \frac{h(\bar t,\bar \eta_{\st})}{h(\bar s,\bar \eta_{\st} )} K_a(\bar x|\bar\eta_{\so}).
 \ee
Here the sum is taken over partitions $\{\bar t,\bar y+c\}=\bar\eta\Rightarrow\{\bar\eta_{\so},\bar\eta_{\st}\}$  such that $\#\bar\eta_{\so}=a$ and $\#\bar\eta_{\st}=b$.

\item As a sum over partitions of $\bar t$ and $\bar x$:
 \begin{multline}\label{GF1}
 Z_{a,b}(\bar t;\bar x|\bar s;\bar y)=
 (-1)^{a}  h(\bar x, \bar x) h(\bar s, \bar x) g(\bar x,\bar y) g(\bar s,\bar y) \\
  \sum K_{a}(\bar t - c |\bar \xi_{\so}) \frac{h(\bar \xi_{\so}, \bar t)
  g(\bar x, \bar \xi_{\so}) g(\bar s, \bar \xi_{\so})}{g(\bar \xi_{\so}, \bar y)} g(\bar \xi_{\so},\bar \xi_{\st}).
 \end{multline}
Here the sum is taken over partitions $\{\bar t,\bar x-c\}=\bar\xi\Rightarrow\{\bxi_{\so},\bxi_{\st}\}$ such that
$\#\bar\xi_{\so}=\#\bar\xi_{\st}=a$.

\item As a sum over partitions of $\bar s$ and $\bar y$:
 \be{S-GF1}
  Z_{a,b}(\bar t;\bar x|\bar s;\bar y)=(-1)^{a+b} f(\bar x,\bar t) f(\bar s,\bar t)  \sum g(\bar \nu_{\so},\bar \nu_{\st})
  K_{a+b}(\{\bar \nu_{\so}, \bar t - c\}|\{\bar x,\bar s\})
 \ee
Here the sum is taken over partitions $\{\bar s-c,\bar y\}=\bar\nu\Rightarrow\{\bar\nu_{\so},\bar\nu_{\st}\}$ such that
$\#\bar\nu_{\so}=\#\bar\nu_{\st}=b$.

%
% \be{S-GF2}
%  Z_{a,b}(\bar t;\bar x|\bar s;\bar y)=(-1)^{a+b} f(\bar x,\bar t) f(\bar s,\bar t)  \sum g(\bar \xi_{\so},\bar \xi_{\st})
%K_{a+b}(\bar y, \bar t - c|\bar  x,\bar \xi_{\st}+c)
% \ee
%

%Here $\bar\xi=\{\bar y ,\bar s - c\}$. The sum is taken with respect to partitions of the set $\bar\xi$ into
%subsets $\bar\xi_{\so}$ and $\bar\xi_{\st}$ with $\#\bar\xi_{\so}=b$ and $\#\bar\xi_{\st}=b$.

\end{itemize}

All the sum formulas listed above follow from \eqref{RHC-IHC} and can be proved via reduction of the sums over partitions to multiple contour
integrals of Cauchy type. Let us show how this method works.

Consider a $b$-fold integral
 \be{Int-Or-for}
 %Z_{a,b}(\bar t;\bar x|\bar s;\bar y)
 \mathcal{I}=\frac{(-1)^{b}}{(2\pi ic)^b b!}\oint\limits_{\bar w} K_{a+b}(\bar w|\{\bar t, \bar z + c\})
 h(\bar z, \bar x) \frac{ g(\bar z, \bar y) g(\bar z, \bar w)}{\Delta_b(\bar z) \Delta'_b(\bar z)}
   \,d\bar z,
   \ee
where $\bar w=\{\bar s,\bar x\}$ and $d\bar z=dz_1,\dots,dz_b$. We have used a subscript $\bar w$ on  the integral symbol in order
to stress that the integration contour for every $z_j$ surrounds the set $\bw$ in
the anticlockwise direction. We also assume that the integration contours do not contain any
other singularities of the integrand. Similar prescription will be kept for all other integral representations
considered below.

The only poles of
the integrand within the integration contours are the points $z_j=w_k$. Evaluating the integral
by the residues in these poles we obtain (see appendix~\ref{A-SumInt} for details)
 \be{Int-Or-for-twin-res}
\mathcal{I}=(-1)^b\sum K_{a+b}(\bar w|\{\bar t,\bar w_{\st}+c\}) h(\bar w_{\st},\bar x) g(\bar w_{\st},\bar y) g(\bar w_{\st},\bar w_{\so}),
   \ee
where the sum is taken over partitions of $\bar w$ into subsets $\bar w_{\so}$ and $\bar w_{\st}$ with
$\#\bar w_{\so}=a$ and $\#\bar w_{\st}=b$. Due to
\eqref{K-K} we have
\be{red-Kab}
K_{a+b}(\bar w|\{\bar t,\bar w_{\st}+c\})=(-1)^b K_{a}(\bar w_{\so}|\bar t),
\ee
and comparing the obtained sum with \eqref{RHC-IHC} we immediately obtain $\mathcal{I}=Z_{a,b}(\bar t;\bar x|\bar s;\bar y)$.

Similarly, one can check that the sum over partitions in \eqref{RHC-IHC} can be presented as an
$a$-fold contour integral
 \be{Int-Or-for-twin}
 Z_{a,b}(\bar t;\bar x|\bar s;\bar y)=(-1)^b\frac{h(\bar w,\bar x)g(\bar y,\bar w)}{(2\pi ic)^a a!}\oint\limits_{\bar w}
  \frac{K_a(\bar z|\bar t) g(\bar z,\bar w) }{h(\bar z,\bar x)g(\bar z, \bar y)\Delta_a(\bar z) \Delta'_a(\bar z)}
  \,d\bar z,
   \ee
where now $d\bar z=dz_1,\dots,dz_a$. Indeed, taking the residues in the points $\bz=\bw_{\so}$ we obtain
 \be{Int-AOr-for-twin}
 Z_{a,b}(\bar t;\bar x|\bar s;\bar y)=(-1)^b h(\bar w,\bar x)g(\bar y,\bar w)\sum
  \frac{K_a(\bw_{\so}|\bar t) g(\bw_{\so},\bw_{\st}) }{h(\bw_{\so},\bar x)g(\bw_{\so}, \bar y)}.
   \ee
Multiplying the terms of the sum with the prefactor $h(\bar w,\bar x)g(\bar y,\bar w)$ we arrive at \eqref{RHC-IHC}.

Let us turn back to the integral \eqref{Int-Or-for}. Obviously, it can be calculated taking the residues in
the poles outside the original integration contour. It
is easy to see that for arbitrary $z_j$ the integrand  behaves as $1/z_j^3$ at $z_j\to\infty$. Hence, the
residue at infinity vanishes. The  poles outside the original integration contours are in  $z_j=y_k$ and $z_j=s_k-c$
(the poles at $z_j=x_k-c$ are compensated by the zeros of the product $h(\bar z, \bar x)$). Thus, we can move the original
contour surrounding $\bar w$ to the points $\bar \nu=\{\bar y,\bar s-c\}$
 \be{Int-Or-for-twin1}
 Z_{a,b}(\bar t;\bar x|\bar s;\bar y)=\frac{1}{(2\pi ic)^b b!}\oint\limits_{\bar \nu} K_{a+b}(\bar w|\{\bar t, \bar z + c\})
 h(\bar z, \bar x)\frac{ g(\bar z, \bar y) g(\bar z, \bar w)}{\Delta_b(\bar z) \Delta'_b(\bar z)}
   \,d\bar z.
   \ee
It is convenient  to transform  the integrand, applying \eqref{Red-K} to
$K_{a+b}(\bar w|\{\bar t, \bar z + c\})$. Then substituting $\bar w = \{\bar x, \bar s\}$ and using elementary properties of $f(z,w)$ we obtain
 \be{Int-Or-for-twin2}
 Z_{a,b}(\bar t;\bar x|\bar s;\bar y)=\frac{(-1)^{a+b}}{(2\pi ic)^b b!}\oint\limits_{\bar \nu}
 \frac{K_{a+b}(\{\bar t - c, \bar z\}| \bar w)f(\bar w,\bar t) h(\bar z, \bar x)g(\bar z, \bar y) g(\bar z, \bar x) g(\bar z, \bar s)}
 {f(\bar z, \bar x) f(\bar z,\bar s)\Delta_b(\bar z) \Delta'_b(\bar z)}
   \,d\bar z,
   \ee
and after simplification we arrive at
 \be{Int-Or-for-twin3}
 Z_{a,b}(\bar t;\bar x|\bar s;\bar y)=\frac{(-1)^{a+b} f(\bar w,\bar t)}{(2\pi ic)^b b!}\oint\limits_{\bar \nu}K_{a+b}(\{\bar t - c, \bar z\}| \bar w)
 \frac{g(\bar z, \bar \nu)}{\Delta_b(\bar z) \Delta'_b(\bar z)}
   \,d\bar z.
   \ee
Now all the poles are explicitly combined in the product $g(\bar z,\bar\nu)$. Hence, the
result of the integration gives the sum over partitions of $\bar\nu\Rightarrow\{\bar\nu_{\so},\bar\nu_{\st}\}$ with
$\#\bar\nu_{\so}= \# \bar\nu_{\st}=b$, which coincides with \eqref{S-GF1}.

Applying \eqref{K-K} to the DWPF $ K_{a+b}(\{\bar \nu_{\so}, \bar t - c\}|\{\bar x,\bar s\})$ in \eqref{S-GF1}, we have
\begin{multline}
 K_{a+b}(\{\bar \nu_{\so}, \bar t - c\}|\{\bar x,\bar s\})= (-1)^bK_{a+2b}(\{\bar \nu, \bar t - c\}|\{\bar x,\bar s,\bar \nu_{\st}+c\})\\
 =(-1)^bK_{a+2b}(\{\bar y, \bar s-c,\bar t - c\}|\{\bar x,\bar s,\bar \nu_{\st}+c\})
 =  K_{a+b}(\{\bar y, \bar t - c\}|\{\bar x,\bar \nu_{\st} + c\}).
\end{multline}
Then the sum over partitions in \eqref{S-GF1} is equivalent to a multiple contour integral
 \be{Int-Or-for-twin4}
 Z_{a,b}(\bar t;\bar x|\bar s;\bar y)=\frac{(-1)^{a} f(\bar x,\bar t)f(\bar s,\bar t)}{(2\pi ic)^b b!}\oint\limits_{\bar \nu}
 K_{a+b}(\{\bar y,\bar t - c\}| \{\bar x, \bar z+c\})
 \frac{g(\bar z, \bar \nu)}{\Delta_b(\bar z) \Delta'_b(\bar z)}
   \,d\bar z.
\ee
Using \eqref{Red-K} we recast \eqref{Int-Or-for-twin4} as
 \be{Int-Or-for-twin5}
 Z_{a,b}(\bar t;\bar x|\bar s;\bar y)=\frac{(-1)^{b}f(\bar y,\bar x) f(\bar s,\bar t)}{(2\pi ic)^b b!}\oint\limits_{\bar \nu}
 \frac{K_{a+b}(\{\bar x-c, \bar z\}|\{\bar y,\bar t - c\})}{f(\bar z,\bar y)f(\bar z,\bar t-c)}
 \frac{g(\bar z, \bar \nu)}{\Delta_b(\bar z) \Delta'_b(\bar z)}
   \,d\bar z.
\ee
Setting now $\bet=\{\bt,\by+c\}$ we obtain
 \be{Int-Or-for-twin6}
 Z_{a,b}(\bar t;\bar x|\bar s;\bar y)=\frac{(-1)^{b}f(\bar y,\bar x) f(\bar s,\bar t)}{(2\pi ic)^b b!}\oint\limits_{\bar \nu}
 K_{a+b}(\{\bar x-c, \bar z\}|\bet-c)\frac{h(\bz,\bt)g(\bar z, \bet-2c)}{h(\bz,\bs)\Delta_b(\bar z) \Delta'_b(\bar z)}
   \,d\bar z.
\ee
Now we can evaluate the integral by the residues outside the integration contours. All of them are collected in the
product $g(\bar z, \bet-2c)$. Taking the residues in the points $\bet-2c$ and using $h(x-2c,y)=-h(y,x)$ we immediately
arrive at \eqref{Al-RHC-IHC}.

Similarly, starting with the integral representation \eqref{Int-Or-for-twin} one can obtain the sum formula \eqref{GF1}.

\section*{Conclusion}

In this paper we have derived a sum formula for the scalar product of Bethe vectors in the models with $\mathfrak{gl}(2|1)$
symmetry. We considered the case of generic Bethe vectors. This means that the Bethe parameters are generic complex numbers.
However, one can also use this sum formula, if the Bethe parameters obey some  constraints. In some of these particular cases the
sums over partitions can be taken explicitly, leading eventually to determinant representations for scalar products. In the second
part of this paper we will consider these particular cases in more details. We will show that if a part of Bethe parameters satisfy
Bethe equations, then the sum formula in the $\mathfrak{gl}(2|1)$-based models reduces to a single determinant.

To conclude this paper we would like to mention that our results can be applied to the models with $\mathfrak{gl}(1|2)$
symmetry as well. Indeed, due to an isomorphism between Yangians $Y\bigl(\mathfrak{gl}(1|2)\bigr)$  and
$Y\bigl(\mathfrak{gl}(2|1)\bigr)$ (see \cite{PakRS16a}), it is enough to make the replacements $\{\buc,\bub\}
\leftrightarrow \{\bvc,\bvb\}$ and $a\leftrightarrow b$ in the sum formula. The obtained expression describes the
scalar product of Bethe vectors in $\mathfrak{gl}(1|2)$-based models.

\section*{Acknowledgements}
The work of A.L.  has been funded by the  Russian Academic Excellence Project 5-100 and by joint NASU-CNRS project F14-2016.
The work of S.P. was supported in part by  the RFBR grant 14-01-00474 and  the grant
of  the Scientific Foundation of NRU HSE.
N.A.S. was  supported by  the grants RFBR-15-31-20484-mol-a-ved and RFBR-14-01-00860-a.

%%%%%%%%%%%%%%%%%%%%%%%%%%%%%%%%

\appendix

\section{Properties of  DWPF\label{A-IHC}}

The DWPF  $K_n(\bx|\by)$ is a symmetric function of $x_1,\dots,x_n$ and symmetric function of $y_1,\dots,y_n$.
It behaves as $1/x_n$ (resp. $1/y_n$) as $x_n\to\infty$ (resp. $y_n\to\infty$) at other variables fixed.
It has simple poles at $x_j=y_k$. It follows directly from the definition \eqref{K-def} that  DWPF possesses the following properties:
 \be{K-K}
K_{n+m}(\{\bar x, \bz-c\}|\{\bar y, \bz\})=K_{n+m}(\{\bar x, \bz\}|\{\bar y, \bz+c\})= (-1)^m K_{n}(\bar x|\bar y), \qquad
\#\bz=m,
\ee
and
 \be{Red-K}
K_{n}(\bar x-c|\bar y)=K_{n}(\bar x|\bar y+c)= (-1)^n \frac{ K_{n}(\bar y|\bar x)}{f(\bar y,\bar x)}.
\ee

\section{Sum over partitions as a contour integral\label{A-SumInt}}

\begin{prop}\label{P-SumInt}
Let $\bw=\{w_1,\dots,w_N\}$ be a set of complex numbers. Let $\mathcal{F}(\bz)$ be a function of $n$ variables
$z_1,\dots,z_n$ ($n\le N$). Assume that $\mathcal{F}(\bz)$ is a symmetric function of $\bz$ and that
it is holomorphic with respect to
each $z_j$ within a domain containing the points $\bw$. Define
\be{F-Int}
\langle\mathcal{F}\rangle=\frac1{(2\pi i c)^nn!}\oint\limits_{\bw}\frac{g(\bz,\bw)\,d\bz}{\Delta_n(\bz)\Delta'_n(\bz)}\mathcal{F}(\bz).
\ee
Here $d\bz=dz_1,\dots,dz_n$ and the integration contour for every $z_j$ surrounds the points $\bw$ in the anticlockwise direction.
We assume that there is no other singularities of the integrand within the integration contours.
Then
\be{F-Sum}
\langle\mathcal{F}\rangle=\sum  g(\bw_{\so},\bw_{\st})\mathcal{F}(\bw_{\so}),
\ee
where the sum is taken over partitions $\bw\Rightarrow\{\bw_{\so},\bw_{\st}\}$ such that $\#\bw_{\so}=n$.
\end{prop}

{\sl Proof.}
We use induction over $n$. For $n=1$ the statement of the proposition is obvious. Suppose that it is valid for some $n-1$.
Then splitting $\bz=\{z_n,\bz_n\}$ we obtain
\begin{multline}\label{Sum-Int}
\langle\mathcal{F}\rangle=\frac1{(2\pi i c)^nn!}\oint\limits_{\bw}\frac{g(z_n,\bw)g(\bz_n,\bw)\mathcal{F}(\{\bz_n,z_n\})\,d\bz_n\,dz_n}
{\Delta_{n-1}(\bz_n)\Delta'_{n-1}(\bz_n)g(z_n,\bz_n)g(\bz_n,z_n)}\\
=\sum \frac{g(\bw_{\so},\bw_{\st})}{2\pi i cn}\oint\limits_{\bw}\frac{g(z_n,\bw_{\st})\mathcal{F}(\{\bw_{\so},z_n\})\,dz_n}
{g(\bw_{\so},z_n)},
\end{multline}
where the sum is taken over partitions $\bw\Rightarrow\{\bw_{\so},\bw_{\st}\}$ such that $\#\bw_{\so}=n-1$. Performing the integration
over $z_n$ we find
\be{2-part}
\langle\mathcal{F}\rangle=\frac1n\sum \frac{g(\bw_{\so},\bw_{\st})g(\bw_{\qo},\bw_{\qt})}
{g(\bw_{\so},\bw_{\qo})}\mathcal{F}(\{\bw_{\so},\bw_{\qo}\}),
\ee
where we obtain an additional partition $\bw_{\st}\Rightarrow\{\bw_{\qo},\bw_{\qt}\}$ with $\#\bw_{\qo}=1$.
Substituting in \eqref{2-part}
$\bw_{\st}=\{\bw_{\qo},\bw_{\qt}\}$  and setting there  $\{\bw_{\qo},\bw_{\so}\}=\bw_{0}$ we arrive at
\be{F-Sum1}
\langle\mathcal{F}\rangle=\frac1n\sum  g(\bw_{0},\bw_{\qt})\mathcal{F}(\bw_{0}).
\ee
Now the sum over partitions is organized as follows. First we have the partitions $\bw\Rightarrow\{\bw_{0},\bw_{\qt}\}$ with
$\#\bw_{0}=n$, and then we have the additional partition $\bw_{0}\Rightarrow\{\bw_{\qo},\bw_{\so}\}$ with $\#\bw_{\qo}=1$. Obviously,
the sum over the later partition gives $n$, and we obtain the statement of the proposition.
\qed

Note that if $n>N$ in \eqref{F-Int}, then $\langle\mathcal{F}\rangle=0$.

\section{Summation formulas \label{A-SF}}

\begin{lemma}\label{main-ident}
Let $\bar\xi$, $\bar\alpha$ and $\bar\beta$ be sets of complex variables with $\#\alpha=n$,
$\#\beta=m$, and $\#\xi=n+m$. Then
\begin{equation}\label{Sym-Part-old1}
  \sum
 K_{n}(\bar\xi_{\so}|\bar \alpha)K_{m}(\bar \beta|\bar\xi_{\st})f(\bar\xi_{\st},\bar\xi_{\so})
 = (-1)^{n}f(\bar\xi,\bar \alpha) K_{n+m}(\bar \alpha-c,\bar \beta|\bar\xi).
 \end{equation}
The sum is taken with respect to all partitions of the set $\bar\xi$ into
subsets $\bar\xi_{\so}$ and $\bar\xi_{\st}$ with $\#\bar\xi_{\so}=n$ and $\#\bar\xi_{\st}=m$.
\end{lemma}

The proof of this lemma can be found in \cite{BelPRS12a}.

\begin{lemma}
For any set of functions $\phi_k(\beta)$, $k=1,\dots,n+m$, let
\be{Detphi}
\Phi_{n+m}(\bbet)=\Delta_{n+m}(\bbet)\det_{n+m}\phi_k(\beta_j),
\ee
where $\bbet=\{\beta_1,\dots,\beta_{n+m}\}$.
Then
\begin{multline}\label{Detphi1}
\Phi_{n+m}(\bbet)=\sum \Delta_{n}(\bbet_{\so})\det_{k=1,\dots,n}\phi_k(\beta_{{\so}_j})\cdot
\Delta_{m}(\bbet_{\st})\det_{k=1,\dots,m}\phi_{n+k}(\beta_{{\st}_j})\cdot g(\bbet_{\st},\bbet_{\so})\\
=\sum \Phi_n(\bbet_{\so}) \wh\Phi_m(\bbet_{\st}) g(\bbet_{\st},\bbet_{\so}),
\end{multline}
where $\Phi_n(\bxi_{\so})$ is built on the functions $\phi_k$, $k=1,\dots,n$, while
$\wh\Phi_m(\bxi_{\st})$ is built on the functions $\phi_{n+k}$, $k=1,\dots,m$.
\end{lemma}
\proof
Developing the determinant in \eqref{Detphi} over the first $n$ columns via Laplace formula
we obtain
\be{Detphi0}
\Phi_{n+m}(\bbet)=\Delta_{n+m}(\bbet)\sum (-1)^{P_{\so,\st}}\det_{k=1,\dots,n}\phi_k(\beta_{{\so}_j})\det_{k=1,\dots,m}\phi_{n+k}(\beta_{{\st}_j}),
\ee
where the sum is taken over partitions $\bbet\Rightarrow\{\bbet_{\so},\bbet_{\st}\}$ such that $\#\bbet_{\so}=n$.
The sign $P_{\so,\st}$ is the parity of a permutation mapping the union $\{\bbet_{\so},\bbet_{\st}\}$ into the naturally ordered
set $\bbet$. One can get rid of this sign presenting $\Delta_{n+m}(\bbet)$ as follows
\be{Del-sig}
\Delta_{n+m}(\bbet)=(-1)^{P_{\so,\st}}\Delta_{n}(\bbet_{\so})\Delta_{m}(\bbet_{\st}) g(\bbet_{\st},\bbet_{\so}).
\ee
 Substituting \eqref{Del-sig} into \eqref{Detphi0} we immediately arrive at \eqref{Detphi1}. 
\qed

We use several particular cases of \eqref{Detphi1} in the core of the paper. Let
\be{matelJ-1}
\begin{aligned}
&\phi_{k}(\beta)=\frac{g(\beta,x_k)}{h(\beta,x_k)},&\qquad k=1,\dots,n;\\
&\phi_{k+n}(\beta)=g(\beta,y_k)\frac{h(\beta,\bar t)}{h(\beta,\bar s)},&\qquad k=1,\dots,m,
\end{aligned}
\ee
where $\bx$, $\by$, $\bt$, and $\bs$ are some sets of parameters. Then the matrix elements $\phi_{k}(\beta_j)$ coincide
with the entries $\mathcal{J}_{jk}$ \eqref{matelJ}. Hence, we obtain for $J_{n,m}(\bar x;\bar y|\bar t;\bar s|\bar\beta)$
\eqref{def-J}
\begin{multline}\label{J-sumpar}
J_{n,m}(\bar x;\bar y|\bar t;\bar s|\bar\beta)=
\Delta'_n(\bar x)\Delta'_m(\bar y)\;
\sum \Delta_n(\bbet_{\so})\det_n\left(\frac{g(\beta_{{\so}_j},x_k)}{h(\beta_{{\so}_j},x_k)}\right)\\
\times \Delta_m(\bbet_{\st})\det_m\left(g(\beta_{{\st}_j},y_k)
\frac{h(\beta_{{\st}_j},\bar t)}{h(\beta_{{\st}_j},\bar s)}\right)g(\bbet_{\st},\bbet_{\so}).
\end{multline}
Now we use the definition of DWPF
\be{KK-def}
\Delta'_n(\bar x)\Delta_n(\bbet_{\so})\det_n\left(\frac{g(\beta_{{\so}_j},x_k)}{h(\beta_{{\so}_j},x_k)}\right)=\frac{K_n(\bbet_{\so}|x)}
{h(\bbet_{\so},\bx)},
\ee
and an explicit expression \eqref{Cauchy-0} for Cauchy determinant $\det_m g(\beta_{{\st}_j},y_k)$.
Substituting these expressions into \eqref{J-sumpar} we find
\begin{equation}\label{J-sumpar-1}
J_{n,m}(\bar x;\bar y|\bar t;\bar s|\bar\beta)=
\sum \frac{K_n(\bbet_{\so}|x)}{h(\bbet_{\so},\bx)}
\cdot g(\bbet_{\st},\by) \frac{h(\beta_{\st},\bar t)}{h(\beta_{\st},\bar s)}g(\bbet_{\st},\bbet_{\so}),
\end{equation}
which coincides with \eqref{J-sum}.

Another example used in the text is
\be{twoCauchy}
\begin{aligned}
&\phi_{k}(\beta)=g(\beta,x_k),&\qquad k=1,\dots,n;\\
&\phi_{k+n}(\beta)=g(\beta,y_k),&\qquad k=1,\dots,m.
\end{aligned}
\ee
Then using explicit representation of the Cauchy determinant \eqref{Cauchy-0} we have
\be{PhiCau}
\Phi_{n+m}(\bbet)=\frac{g(\bbet,\bx)g(\bbet,\by)}{\Delta'_{n+m}(\{\bx,\by\})}.
\ee
On the other hand, it follows from \eqref{Detphi1} that
\begin{equation}\label{Detphi2}
\frac{g(\bbet,\bx)g(\bbet,\by)}{\Delta'_{n+m}(\{\bx,\by\})}=\sum \Delta_{n}(\bbet_{\so})\det_{k=1,\dots,n}g(\beta_{{\so}_j},x_k)\cdot
\Delta_{m}(\bbet_{\st})\det_{k=1,\dots,m}g(\beta_{{\st}_j},y_k)\cdot g(\bbet_{\st},\bbet_{\so}).
\end{equation}
Multiplying \eqref{Detphi2} with $\Delta'_n(\bx)$ and $\Delta'_m(\by)$ and using \eqref{Cauchy-0} we arrive at

\begin{equation}\label{Sym-Part-new1}
  \sum g(\beta_{\so},\bx) g(\beta_{\st},\by) g(\bbet_{\st},\bbet_{\so})
=\frac{g(\bbet,\bx)g(\bbet,\by)}{g(\bx,\by)},
 \end{equation}
where the sum is taken with respect to the partitions of the set $\bbet$ into
subsets $\bbet_{\so}$ and $\bbet_{\st}$ with $\#\bbet_{\so}=n$ and $\#\bbet_{\st}=m$.

\section{Reduction properties of $J_{n,m}$\label{A-RPJ}}

Consider a function $J_{n+1,m}(\{\bar x,z'\};\bar y|\bar t;\bar s|\{\bar\beta,z\})$ defined by \eqref{def-J}.
Let $z'\to z$. Then the matrix element $g(z,z')/h(z,z')$ becomes singular and the determinant reduces to the
product of this singular element and the corresponding minor. After elementary algebra we obtain
\be{limW-0}
\lim_{z'\to z}\frac1{g(z,z')} J_{n+1,m}(\{\bar x,z'\};\bar y|\bar t;\bar s|\{\bar\beta,z\})=
g(\bar \beta,z)g(z,\bar x)J_{n,m}(\bar x;\bar y|\bar t;\bar s|\bar\beta).
\ee

Similarly, if we consider the function $J_{n,m+1}(\bar x;\{\bar y,z'\}|\bar t;\bar s|\{\bar\beta,z\})$ in the
limit $z'\to z$, then the matrix element $g(z,z')h(z,\bar t)/h(z,\bar s)$ becomes singular. The determinant
again reduces to the product of this singular element and the corresponding minor, and we find
\be{limW-02}
\lim_{z'\to z}\frac1{g(z,z')} J_{n,m+1}(\bar x;\{\bar y,z'\}|\bar t;\bar s|\{\bar\beta, z\})=
g(z,\bar \beta)g(\bar y,z)\frac{h(z,\bar t)}{h(z,\bar s)}\;J_{n,m}(\bar x;\bar y|\bar t;\bar s|\bar\beta).
\ee

Equations \eqref{limW-0}, \eqref{limW-02} obviously could be generalized to the case when $z$ and $z'$ are respectively replaced with
the sets $\bz$ and $\bz'$ such that
$\#\bar z=\#\bar z'=\rho\ge 1$. Then
\be{limW-1}
\lim_{\bar z'\to\bar z}\frac1{g(\bar z,\bar z')} J_{n+\rho,m}(\{\bar x,\bar z'\};\bar y|\bar t;\bar s|\{\bar\beta,\bar z\})=
g(\bar \beta,\bar z)g(\bar z,\bar x)J_{n,m}(\bar x;\bar y|\bar t;\bar s|\bar\beta),
\ee
and
\be{limW-2}
\lim_{\bar z'\to\bar z}\frac1{g(\bar z,\bar z')} J_{n,m+\rho}(\bar x;\{\bar y,\bar z'\}|\bar t;\bar s|\{\bar\beta,\bar z\})=
g(\bar z,\bar \beta)g(\bar y,\bar z)\frac{h(\bar z,\bar t)}{h(\bar z,\bar s)}\;J_{n,m}(\bar x;\bar y|\bar t;\bar s|\bar\beta).
\ee
One more obvious reduction is
\be{sim-W}
J_{n,m}(\bar x;\bar y|\{\bar t,\bar z\};\{\bar s,\bar z\}|\bar\beta)=
J_{n,m}(\bar x;\bar y|\bar t;\bar s|\bar\beta).
\ee
%

%%%%%%%%%%%%%%%%%%%%%%%%%%%%%%%%%%%%%%%%%%%%%%%%%%%%%%%%%%%%%%%%%%%%%%%%%%%%%%

\end{document}